\documentclass[prl,twocolumn,superscriptaddress,showpacs,floatfix,longbibliography]{revtex4-2}
\usepackage{graphicx,subfigure}
\usepackage{epsfig}
\usepackage{bm}				
\usepackage{dcolumn}
\usepackage{color,xcolor}
\usepackage{physics}
\usepackage{float}
\usepackage[colorlinks,urlcolor=cyan,citecolor=blue,linkcolor=magenta]{hyperref}
\makeatletter
\newcommand*{\rom}[1]{\expandafter\@slowromancap\romannumeral #1@}
\makeatother
\usepackage{lipsum}
\usepackage{subfigure}
\usepackage{ulem}
\usepackage{dsfont}
\usepackage{enumitem}
\usepackage{tikz,verbatim}
\usetikzlibrary{quantikz}

\newcommand{\rz}[1]{\textcolor[rgb]{0,0.0,.0}{#1}}
\begin{document}

\title{Observation of the non-Hermitian skin effect and Fermi skin on a digital quantum computer}
\author{Ruizhe Shen}
\email{e0554228@u.nus.edu}
\affiliation{Department of Physics, National University of Singapore, Singapore 117542}

\author{Tianqi Chen}
\email{tianqi.chen@ntu.edu.sg}
\affiliation{School of Physical and Mathematical Sciences, Nanyang Technological University, Singapore 639798}

\author{Bo Yang}
\affiliation{School of Physical and Mathematical Sciences, Nanyang Technological University, Singapore 639798}
\author{Ching Hua Lee}
\email{phylch@nus.edu.sg}
\affiliation{Department of Physics, National University of Singapore, Singapore 117542}
%\affiliation{Joint School of National University of Singapore and Tianjin University, International Campus of Tianjin University, Binhai New City, Fuzhou 350207, China}
\date{\today}

\maketitle

%%%%%%%%%%%%%%%%%%%%%%%%%%%%%%%%%%%%%%%%%%%%%%%%%%%%%%%%%%%%%%%%%%%%%%%%%%%%%%%%%%%%%%%%%%%%%%%%%%%%%%%%%%%%%%
%{\color{blue}\emph{Introduction}}.\textemdash
%\onecolumngrid

%{\color{blue}\noindent{\it Introduction.--}}

%===================================================Introduction=============================================

%{\bf Introduction}
%\section{\bf  Introduction}
{\bf Non-Hermitian physics has attracted considerable attention in recent years, particularly the non-Hermitian skin effect (NHSE) for its extreme sensitivity and non-locality. While the NHSE has been physically observed in various classical metamaterials and even ultracold atomic arrays, its highly-nontrivial implications in many-body dynamics have never been experimentally investigated. In this work, we report the first observation of the NHSE on a universal quantum processor, as well as its characteristic but elusive Fermi skin from many-fermion statistics. To implement NHSE dynamics on a quantum computer, the effective time-evolution circuit not only needs to be non-reciprocal and non-unitary but must also be scaled up to a sufficient number of lattice qubits to achieve spatial non-locality. We show how such a non-unitary operation can be systematically realized by post-selecting multiple ancilla qubits, as demonstrated through two paradigmatic non-reciprocal models on a noisy IBM quantum processor, with clear signatures of asymmetric spatial propagation and many-body Fermi skin accumulation. To minimize errors from inevitable device noise, time evolution is performed using a trainable, optimized quantum circuit produced with variational quantum algorithms. Our study represents a critical milestone in the quantum simulation of non-Hermitian lattice phenomena on present-day quantum computers and can be readily generalized to more sophisticated many-body models with the remarkable programmability of quantum computers. 
}
\\

\noindent{\bf Introduction.} Contemporary research has revealed a variety of intriguing non-Hermitian phenomena~\cite{kawabata2019symmetry,ashida2020non}, significantly expanding the horizon of quantum physics. Notably, the celebrated non-Hermitian skin effect (NHSE) has garnered intense interest, sparked off by challenging established results in topological bulk-boundary correspondences (BBCs)~\cite{yao2018edge,lee2019anatomy,PhysRevLett.116.133903,kunst2018biorthogonal,PhysRevLett.121.026808} and inspiring new notions of spectral topology~\cite{okuma2020topological,zhang2020correspondence,lee2020unraveling, li2020topological}. The most fascinating non-Hermitian phenomena occur in the realm of interacting many-body physics involving correlations and many-body statistics. For instance, in the presence of the NHSE, particle statistics can lead to the manifestation of a real-space ``Fermi surface'' (Fermi skin) and non-Hermitian topological localization in high-dimensional spaces~\cite{lee2021many,shen2022non}. Furthermore, insights from
non-unitary conformal field theories allow for the extrapolation of non-unitary quantum criticality and entanglement properties~\cite{PhysRevResearch.2.033069,tu2022renyi,lee2022exceptional, shen2023proposal}. Additionally, the interplay between many-body interactions and non-unitarity also leads to highly nontrivial consequences on many-body localizations and dynamical quantum phase transitions~\cite{wang2023non}.

Yet, much of the above-mentioned exotic non-Hermitian phenomena has remained experimentally elusive, primarily due to challenges in engineering quantum interactions alongside non-Hermiticity. While the single-body version of the \rz{non-Hermitian physics} has been successfully demonstrated in photonic~\cite{xiao2020non}, electrical~\cite{helbig2020generalized, zou2021observation}, active matter~\cite{ghatak2020observation} and ultracold atomic~\cite{liang2022observation} setups, frontier many-body non-Hermitian phenomena require a versatile quantum platform, and experimental demonstrations have been limited.  Ultracold atomic lattices, engineered using optically-trapped atoms, offer an emerging platform for the investigation of many-body phenomena~\cite{greiner2002quantum, bloch2008many}, with non-Hermiticity effectively realized through laser-induced loss~\cite{ren2022chiral,shen2023proposal}. Nonetheless, the engineering of many non-Hermitian many-body models remains a challenge due to imperfect control of interactions among atoms under loss, as well as limitations in realizing long-ranged interactions~\cite{urban2009observation,gaetan2009observation}. Consequently, only a few distinct non-interacting non-Hermitian phenomena have been demonstrated with ultracold atom systems~\cite{ren2022chiral,liang2022dynamic}, if experimentally at all, suggesting the need for alternative, more versatile solutions.

A possible answer lies in the rapid improvements in universal quantum computing capabilities of late~\cite{preskill2018quantum,lamm2018simulation, smith2019simulating}. Compared with ultracold atomic setups, digital quantum computers made of superconducting quantum circuits provide arguably greater programmability, with their exceptional control over qubit couplings~\cite{lamm2018simulation, smith2019simulating}. Such advancements facilitate the on-demand implementation of sophisticated quantum interactions, thereby offering a promising route toward the versatile engineering of many-body non-Hermitian models.

In this work, taking advantage of these newly available capabilities of universal quantum computers, we demonstrate, for the first time, the realization of the non-Hermitian skin effect (NHSE) and its many-fermion ``Fermi skin'' profile on current noisy intermediate-scale quantum (NISQ) processors. This goes beyond previous experiments that are restricted to preparing ground states via non-unitary imaginary-time evolution~\cite{motta2020determining,mcardle2019variational}.
The NHSE, in particular, has never been realized in a fully programmable quantum platform; previous implementations in other platforms~\cite{helbig2020generalized, zou2021observation,scheibner2020non,sone2020exceptional,liang2022observation} all demonstrated its key signature of directed amplification, but only at the {\it single particle} level. However, we have managed to observe the NHSE also at the {\it many-body} level, where the directed NHSE pumping faces strong competition with Fermi degeneracy pressure (Pauli exclusion) from up to half fermionic filling, characterized by a characteristic Fermi skin profile. Crucially, this is the first reported demonstration of an intrinsically non-Hermitian many-body phenomenon.

In observing the NHSE at the many-body level, significant challenges have to be overcome. A primary challenge is to realize robust non-unitary Hamiltonian evolution using quantum hardware based on fundamentally unitary quantum gates. To this end, several studies have attempted to realize non-unitary dynamics from the Lindblad equation~\cite{han2021experimental}. However, simulating the Lindblad equation requires doubling the system, which results in poor circuit scalability~\cite{hu2020quantum}. Although some measurement-based methods can in principle implement non-unitary operators exactly~\cite{mao2022measurement}, they demand the continuous monitoring of circuits, which can introduce significant errors from mid-circuit measurements~\cite{pino2021demonstration} and thereby limit the circuit depth. \rz{The dilation method has been proposed to simulate non-unitary circuits~\cite{PhysRevLett.101.230404,PhysRevLett.119.190401}, but the scalability of this approach remains a significant challenge.}
Another challenge is posed by the significant level of noise and decoherence in present-day NISQ devices, which limits individual qubit and gate fidelities~\cite{preskill2018quantum} such that reasonable signal-to-noise ratios can only be attained in relatively shallow quantum circuits.

To surmount these challenges, we devise an ancilla-based framework for lattice systems, where generic non-unitary operators are implemented by embedding them within unitary operators involving extra ``ancilla'' qubits. \rz{While ancilla-based approaches have previously been demonstrated on small systems~\cite{kamakari2022digital,xiao2019observation,wu2019observation}, our approach allows for condensed matter lattice simulations for larger system sizes, and it enables greater programmability in implementing the desired non-Hermitian Hamiltonian evolution.} Significantly, the scalability of our framework allows for the implementation of the effectively asymmetric couplings necessary for an extended NHSE lattice, which is a costly endeavor with the existing Pauli string approximation method~\cite{motta2020determining,kamakari2022digital}.  To transcend the circuit complexity limitations placed by noise and decoherence, we also extensively employed readout-error mitigation methods~\cite{mthree} and circuit recompilation~\cite{heya2018variational,koh2022stabilizing}.

%Trotterized dynamics
%======================================================Results==================================================================

\section{Results}

\subsection{Dynamical signatures of the NHSE and its Fermi skin}
The non-Hermitian skin effect (NHSE) is marked by the extensive pumping of all states in the direction of lattice coupling asymmetry, where left/right hoppings are rescaled by the factor $e^{\pm\kappa}$. Accompanied by gain, this directional pumping is extremely robust and fundamentally disrupts the Bloch nature of lattice basis states, leading to profoundly modified topological bulk-boundary correspondences~\cite{helbig2020generalized}, novel edge bursts~\cite{xue2022non}, emergent non-locality and kinked linear responses~\cite{qin2023kinked}, amongst other enigmatic non-Hermitian phenomena~\cite{lee2019anatomy}.
\begin{figure*}
	\centering
	\includegraphics[width=0.9\linewidth]{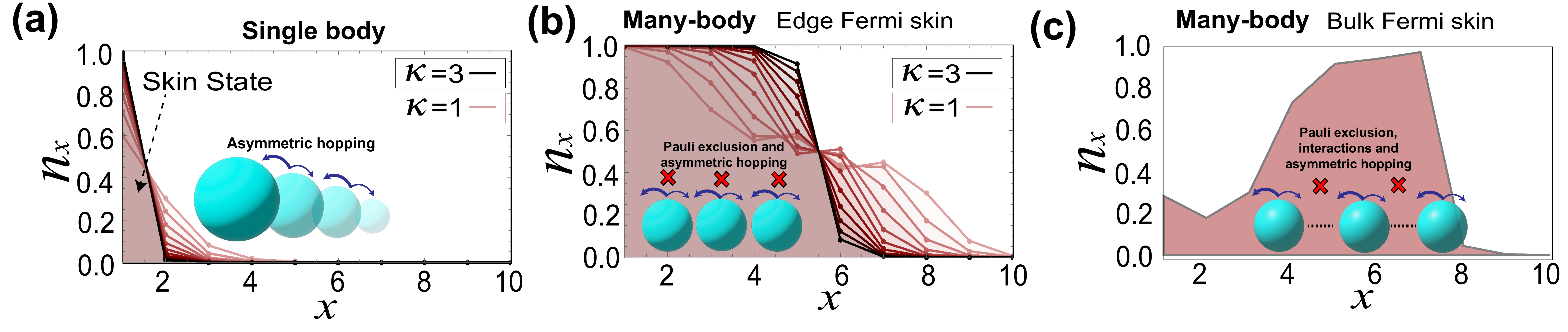}\\
	\caption{{\bf The non-Hermitian skin effect (NHSE) and its many-fermion counterpart.} (a) At the single-particle level, the NHSE from asymmetric lattice hoppings amplifies all states towards one direction, resulting in boundary ``skin'' states with exponential localization lengths inversely proportional to the asymmetry parameter $\kappa$.
	(b) At the many-body level, Pauli exclusion prevents more than one fermion from accumulating at any boundary site, asymptotically resulting in a ``Fermi-skin'' density profile $n_x$ resembling a spatial Fermi-Dirac distribution with effective ``temperature'' proportional to $\kappa$, as plotted with Eq.~\ref{nx0} for the Hatano-Nelson model. \rz{Panel (c) shows the bulk Fermi skin for the interacting  Hatano-Nelson model (Eq.~\ref{int} with $\gamma/J=0.5$ and $U/J=5.0$). The short-range interactions induce skin localization clusters in bulk, which accumulate against each other to form a bulk Fermi skin.} 
}
	\label{fig:nhsemain}
\end{figure*}

For a single particle, this relentless directional pumping amplifies any given wave packet asymmetrically towards one direction as it spreads, such that the wave packet eventually reaches a boundary and accumulates exponentially against it, with a localization ``skin depth'' proportional to $\kappa^{-1}$ (FIG.~\ref{fig:nhsemain}a). This is a universal feature of the NHSE across all platforms both classical or quantum, and has indeed been experimentally demonstrated in lattices made up of electrical circuits~\cite{helbig2020generalized}, mechanical metamaterials~\cite{ghatak2020observation} and ultracold atoms~\cite{liang2022dynamic}. 

What is more interesting, and never experimentally demonstrated, is the NHSE dynamics in the presence of many-body effects \rz{such as Pauli exclusion, even in the absence of interaction terms}. With multiple fermions, Pauli exclusion prevents the NHSE from accumulating all the fermions at one single boundary site, implying that the steady-state density distribution~\cite{mu2020emergent} resembles that of a real-space ``Fermi surface'' with a ``Fermi sea'' at constant density accumulating against a boundary (FIG.f~\ref{fig:nhsemain}b). \rz{In the case without fermion-fermion interaction terms in the Hamiltonian,} this emergent ``Fermi surface'', which we dub the Fermi skin, converges to a Fermi-Dirac-like spatial profile for sufficiently strong non-Hermitian coupling asymmetry i.e. $e^\kappa\gg 1$, and indeed can be characterized by an effective ``temperature'' that is inversely proportional to the strength of $\kappa$ (see Methods). 

As a new conceptual development, the Fermi skin density profile is found to be fundamentally related to the non-orthonormality of the single-particle NHSE eigenstates. Consider a collection of $N$ fermions occupying generic single-body right eigenstates $\{ |\bar\phi_n^R\rangle\}$, $n=1,...,N$ 
that, due to the non-Hermiticity, neither form a complete nor orthonormal basis. Mathematically, $\sum_n\ket{\bar\phi^R_n}\bra{\bar\phi^R_n}\neq 1$ and $B_{mn}=\langle\bar\phi^R_m|\bar\phi^R_n\rangle\neq \delta_{mn}$ respectively, with the overlap matrix $B_{mn}$ keeping track of the non-orthonormality. Even though $B_{mn}$ is defined with the single-particle eigenstates, it directly controls the many-body spatial density via
\begin{equation}
	n_x=\sum_{m,n=1}^N\bar\phi_{m}^{R*}(x)\left[B^{-1}\right]_{mn}\bar\phi_{n}^R(x),
	\label{nx0}
\end{equation}
as derived in the Methods using the Slater determinant of the $N$-fermion state. By diagonalizing $B_{mn}=\sum_\nu b_\nu \varphi_\nu^m\varphi_\nu^n$ into eigenvectors with eigenvalues $b_\nu$ and eigenvectors $\varphi_\nu=(\varphi_\nu^1,...,\varphi_\nu^N)^T$, we can further perform the decomposition $n_x=\sum_\nu n_\nu(x)$ (Eq.~\ref{nxB2}) where
\begin{equation}
	n_\nu(x)=\frac1{b_\nu} \left|\sum_m\varphi_\nu^{m}\bar\phi^R_m(x)\right|^2.
\end{equation}
Since each $\int n_\nu(x)dx=1$ is normalized to unity, $n_x$ has been decomposed into $N$ effective non-overlapping particle density contributions $n_\nu(x)$. Under strong NHSE against a boundary at $x=0$, the first few $n_\nu(x)$ are strongly localized at $x=\nu$. However, as $\nu$ approaches $N$, these $n_\nu(x)$ profiles spread out considerably due to $\{|\bar\phi_n^R\rangle\}$ mixing, leading to a universal Fermi-Dirac-like profile (edge Fermi skin), as plotted in the Methods.

Due to the universality of the Fermi skin profile in the strong NHSE limit, an arbitrary initial state, which would be a superposition of various many-body Slater determinant eigenstates, would generically dynamically evolve to the Fermi skin profile of the eigenstate with the largest $\text{Im}(E)$. In cases with real spectra, as in the models simulated in this work, rapid density fluctuations would indefinitely persist and the Fermi skin profile only emerges in the time-averaged state density. At weaker NHSE strengths or models with nontrivial unit cells, kinks may also slightly punctuate the otherwise Fermi-Dirac-like profile (FIG.~\ref{fig:nhsemain}b).  \rz{Furthermore, this Hamiltonian becomes significantly gapped under the interplay between Pauli exclusion and strong inter-particle interactions (i.e. in Eq.~\ref{int}). This gap can thus strongly isolate clustered fermions which will be aggregated in the bulk of the system without a physical boundary, leading to a new phenomenon known as the bulk Fermi skin~(FIG.~\ref{fig:nhsemain}c). Unlike the previously described edge Fermi skin, such interaction-induced bulk Fermi skins can exhibit density accumulation in a direction opposite from the non-reciprocal coupling, as shown in FIG.~\ref{fig:nhsemain}c.}

\begin{figure*}
	\centering
	%	\begin{tikzpicture}
		%		\node[scale=0.8] {
			%			\begin{quantikz}
				%				&\lstick{\bf $\ket{A}\otimes\ket{A}...$}&\qw&\gate[wires=5]{ \bf U_{imag}(\delta t)}\gategroup[5,steps=1,style={dashed,rounded corners,fill=red!20, inner xsep=2pt},background]{{\sc}}&\text{......}&\gate{{\bf post-selection}}\\
				%				&\lstick{\bf$\ket{\psi}_{1}$}&\gate[wires=4]{\bf U_{real}(\delta t)}\gategroup[4,steps=1,style={dashed,rounded corners,fill=blue!20, inner xsep=2pt},background]{{\sc}}&\qw&\text{......}&\meter{}\\
				%				&\lstick{\bf $\ket{\psi}_{2}$}&\qw&\qw&\text{......}&\meter{}\\
				%				&\lstick{\bf $\ket{\psi}_{3}$}&\qw&\qw&\text{......}&\meter{}\\
				%				&\lstick{\bf $\ket{\psi}_{4}$}&\qw&\qw&\text{......}&\meter{}
				%			\end{quantikz}
			%		};
		%	\end{tikzpicture}
	\includegraphics[width=0.99\linewidth]{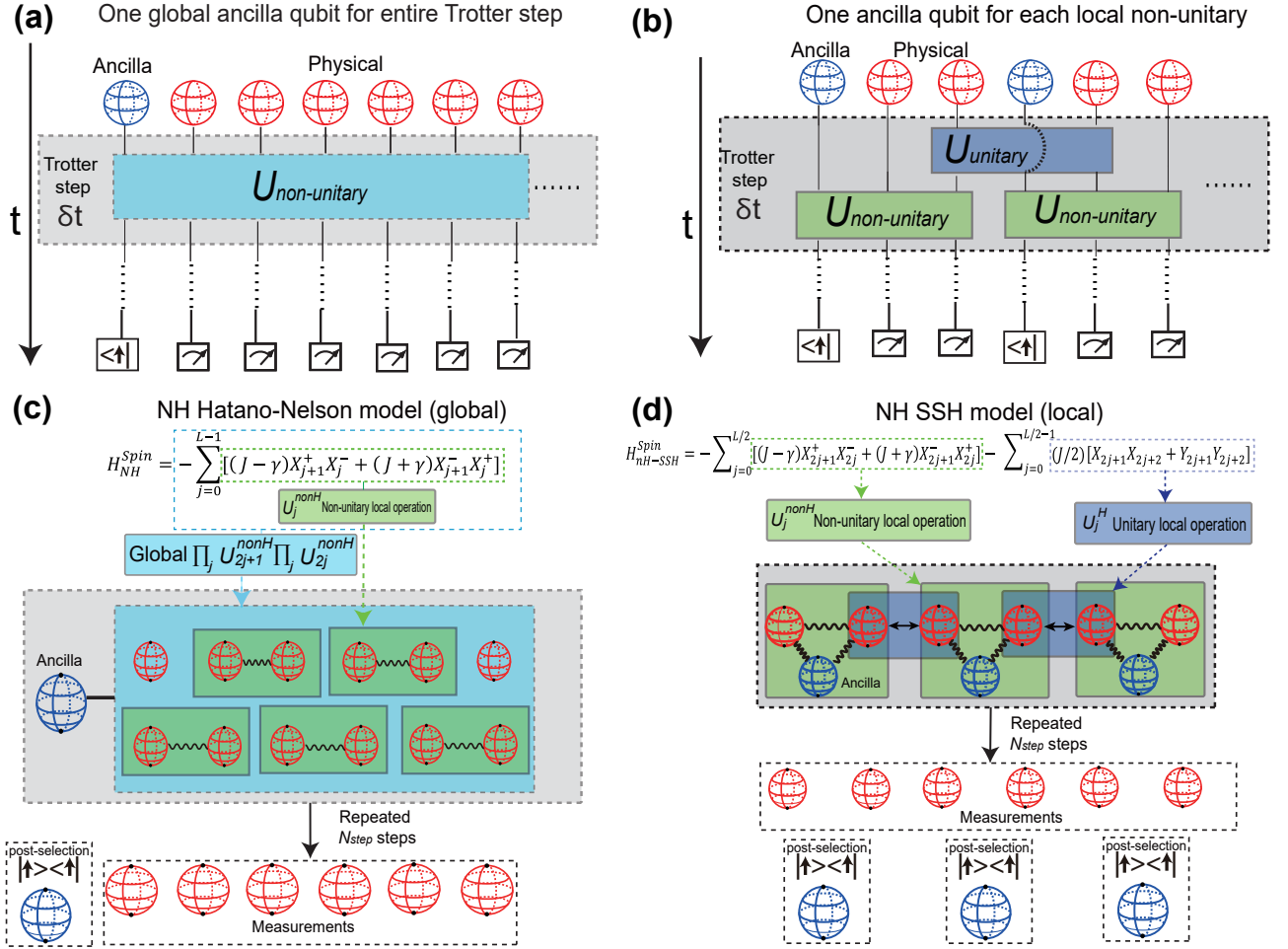}
	\caption{{\bf{Overview of the ancilla-based framework.}}~(a) and (b) show the process of implementing non-unitary dynamics by our local and global ancilla-based circuits respectively. For each Trotter step, 
		the non-unitary process $U_{\rm non-unitary}$ can be realized by coupling N physical qubits (red qubits) with an appropriate number of blue ancilla qubits globally (light blue block) or locally (green block)  (see setups in (c) and (d) respectively).
		In (b), the deep blue block denotes the operator for unitary dynamics, where the ancilla qubits are not coupled.  The post-selection $``\ket{\uparrow}\bra{\uparrow}"$ on all blue ancilla qubits leads to the desired dynamics (see Method). (c) and (d) illustrate concretely how the approaches in (a) and (b) are implemented. (c) shows how the dynamics of the Hatano-Nelson model in Eq. \ref{spinssh_HN} are implemented via the global approach. Here, each green block represents the non-unitary bond $U^{\rm nonH}_{j}=e^{+i\delta t( (J-\gamma)\hat{X}^{+}_{j+1}\!\!\hat{X}^{-}_{j}+(J+\gamma)\hat{X}^{-}_{j+1}\!\!\hat{X}^{+}_{j})}$, which contributes to $(\prod_{j}U^{\rm nonH}_{2j+1})(\prod_{j}U^{\rm nonH}_{2j})$ (light blue) for each Trotter step, which can be realized by coupling with a single blue ancilla qubit (see Methods).	
	(d) shows the circuit with local non-unitary bonds for the non-Hermitian SSH chain in Eq.~\ref{spinssh_main}. For every green block, three waves couple a blue ancilla with two physical qubits (two red qubits), which embed the operator $U^{\rm nonH}_{j}$ in a three-qubit unitary. The double arrows in blue blocks denote the unitary coupling $e^{+iJ\delta t( \hat{X}_{j+1}\!\!\hat{X}_{j}+\hat{Y}_{j+1}\!\!\hat{Y}{_j})}$.  The $\ket{\uparrow}$ outcome of each ancilla qubit from each Trotter step is post-selected after all the $N$ Trotter steps have been performed.
	}
	\label{Hy_main}
\end{figure*}

\subsection{Quantum circuit implementation of non-Hermitian evolution}

In this work, we first demonstrate the usual single-particle NHSE on a quantum computer by dynamically evolving single-fermion initial states with the well-known Hatano-Nelson and non-Hermitian Su-Schrieffer-Heeger (SSH) Hamiltonians. Next, we demonstrate the appearance of a Fermi skin when multiple fermions (spin-up qubits) are similarly evolved.

\subsubsection{General approach: Trotterization and post-selection}
To simulate non-Hermitian evolution with quantum circuits, what needs to be realized is the time evolution operator $e^{-i\hat{H}t}$, which can be effectively implemented from the specified Hamiltonian $\hat H$ by discretizing the evolution into Trotter steps $\delta t$:
\begin{equation}\label{tr}
	\begin{aligned}
		e^{-i\hat{H}t}= [e^{-i\hat{H}\delta t}]^{N_\text{steps}}.
	\end{aligned}
\end{equation}
where $N_\text{steps}=t/\delta t$. This decomposition applies to non-Hermitian models as well, wherein every Trotter step corresponds to a fundamentally non-unitary process. However, each non-unitary step poses a significant obstacle in gate decomposition, since robust non-unitary gates, which can introduce significant decoherence, are still impractical to directly realize on state-of-the-art quantum processors. While one can approximate a non-unitary process in terms of Pauli strings through the quantum imaginary time evolution (QITE) approach~\cite{motta2020determining}, this approximation method is primarily developed as a means to approach ground states, rather than the {\it physical} implementation of non-unitary operators. As discussed, implementing this costly method requires a deep circuit, which can lead to significant errors
on noisy devices.

To encode the non-Hermitian evolution in a quantum circuit as physically as possible, we implement each non-unitary operator by post-selection on a larger unitary operator with extra ``ancilla'' qubits \cite{PhysRevLett.101.230404,PhysRevLett.119.190401,wu2019observation}.  The key idea is that any non-unitary operator $R$ acting on $N$ physical qubits can be embedded in a ($N+1$)-qubit operator $U_{R}$ with an additional ancilla qubit~\cite{lin2021real}. The non-unitary evolution $R$ on a $N$-qubit state $\ket{\psi}$ can be implemented by acting a suitably designed $U_R$ on the $N+1$-qubit state $\ket{\psi}\otimes\ket{\uparrow}_{\text{ancilla}}$, and then projecting the resultant state onto the $\ket{\uparrow}_{\text{ancilla}}$ subspace:
\begin{equation}\label{ur}
	P_{\bra{\uparrow}}[U_{R}(\ket{\psi}\otimes\ket{\uparrow}_{\text{ancilla}})]=R\ket{\psi},
\end{equation}
where $P_{\bra{\uparrow}}$ denotes the projection onto the $\ket{\uparrow}_{\text{ancilla}}$ subspace, executed by post-selecting measurement results with the ancilla qubit pointing upwards. In the $\{\ket{\uparrow},\ket{\downarrow}\}$ basis of the ancilla qubit, this amounts to designing a $2^{N+1}\times 2^{N+1}$ unitary matrix $U_R$ such that its upper left $2^N\times 2^N$ block corresponds to the non-unitary matrix $R$ (see Methods for details on solving for $U_{R}$). 

To construct our quantum circuit for $e^{-i\hat Ht}$, we construct the circuit for each non-unitary Trotter step, and then concatenate the circuits across all time steps. In principle, it is possible to simply realize each non-unitary Trotter step with a single ``global'' ancilla qubit, such that $R=e^{-i\hat{H}\delta t}$ in Eq.~\ref{ur}~(see the global setup in FIG.~\ref{Hy_main}a), as implemented for the Hatano-Nelson model introduced later. While using only a single ancilla qubit indeed minimizes the readout error and discarded information (for $N_A$ ancilla qubits, only approximately $1/2^{N_A}$ of the total shots are post-selected and used), leading to optimal robustness against device noise, it is of limited scalability in practice. This is because of the prohibitive amount of time needed to (classically) solve for the other $2^{N}\times 2^{N}$ blocks of the matrix $U_R$, which is not only costly but also prone to significant numerical errors.

The other alternative would be to approximately decompose each non-unitary step $e^{-i\hat H \delta t}$ into local Hermitian $\hat{H}_{j}^{\rm 0}$ and non-Hermitian $\hat{H}_{j}^{\rm nonH}$ terms, and then use an ancilla qubit for each local non-Hermitian evolution term. With first-order Suzuki-Trotter decomposition~\cite{smith2019simulating} for instance, we have
\begin{equation}\label{hy2}
	\begin{aligned}
		%	e^{-i\hat{H}\delta t}&=e^{-i\hat{H}_{\rm 0}\delta t}e^{-i\hat{H}_{\rm NH}\delta t}+\mathcal{O}(\delta t^2)\\
		e^{-i\hat{H}\delta t}\approx \prod_{j}[e^{-i\hat{H}_{j}^{\rm 0}\delta t}e^{-i\hat{H}_{j}^{\rm nonH}\delta t}].
		%	\nonumber &\approx e^{-i\hat{H}_{\rm 0}\delta t}e^{-i\hat{H}_{\rm NH}\delta t}=U^{\rm H}U^{\rm nonH},
	\end{aligned}
\end{equation}
Here, unitary terms $e^{-i\hat{H}_{j}^{\rm 0}\delta t}$ can be decomposed into a series of basic unitary gates~\cite{lamm2018simulation, smith2019simulating}, and each local non-unitary term is embedded into a larger unitary operator $U_R$ corresponding to $R=e^{-i\hat{H}_{j}^{\rm nonH}\delta t}$, with one ancilla qubit. As sketched in FIG.~\ref{Hy_main}b, the exact unitary and non-unitary operator decomposition depends on the locality structure of the Hermitian and non-Hermitian bonds and is thus model-dependent. The merit of this ``local'' architecture is that one only needs to solve for the $U_R$ for each local non-unitary bond, which induces inconsequential numerical errors due to the very small dimensionality. Moreover, such stacked local unitary and non-unitary operators can be generalized to larger sizes easily, thereby exhibiting good scalability. However, since the number of measurements needed for performing all post-selections naively scales like the exponential of the number of ancilla qubits, this ``local decomposition'' approach may be fraught with significant read errors.

Whether the ``global'' or the ``local decomposition'' approach (or a combination thereof) is more tenable involves a compromise between the acceptable error tolerances and available computational resources/techniques; note that neither have hitherto been physically achieved on a quantum processor in the realization of a non-Hermitian model that is sufficiently large and sophisticated to exhibit the NHSE.

\subsubsection{Implementation of specific models} 
We first consider the Hatano-Nelson model with asymmetric couplings $J\pm \gamma$ between neighboring sites, which is arguably the simplest 1D non-Hermitian lattice model exhibiting the NHSE, with asymmetry ratio parameter $\kappa=\log\sqrt{\frac{J+\gamma}{J-\gamma}}$. While it typically takes the form of a particle-hopping model, for implementation on the IBM quantum processor, we convert it from a fermion model to a spin chain model via the Jordan-Wigner transformation (see the supplement~\cite{supp}):
\begin{equation}\label{spinssh_HN}
	\begin{aligned}
		\hat{H}^{\rm Spin}_{\rm HN}=-\sum_{j=0}^{L-1}\left((J+\gamma)\hat{X}^{+}_{j}\hat{X}^{-}_{j+1}+(J-\gamma)\hat{X}^{-}_{j}\!\hat{X}^{+}_{j+1}\right),
	\end{aligned}
\end{equation}
where each spin-1/2 qubit represents either the presence ($\ket{\uparrow}$) or absence ($\ket{\downarrow}$) of a fermion. For each site $j$, \rz{$\hat{X}^{+}_j=\ket{\uparrow}\!\bra{\downarrow}$ and $\hat{X}^{-}_j=\ket{\downarrow}\!\bra{\uparrow}$ flip the spin-1/2 on that site.}

Since there are no local (on-site) non-Hermitian couplings, and the model is relatively simple overall, we use it to illustrate the global ancilla approach (FIG.~\ref{Hy_main}a).  In this case, we can  straightforwardly set $R=e^{-i{\hat H}^\text{Spin}_\text{HN}\delta t}$ 
to be the global non-unitary operator which is implemented via post-selection on a single ancilla qubit (Eq.~\ref{ur}), for each time step $\delta t$, as sketched in FIG.~\ref{Hy_main} (c).

\begin{figure*}
	\centering
	\includegraphics[width=0.93\linewidth]{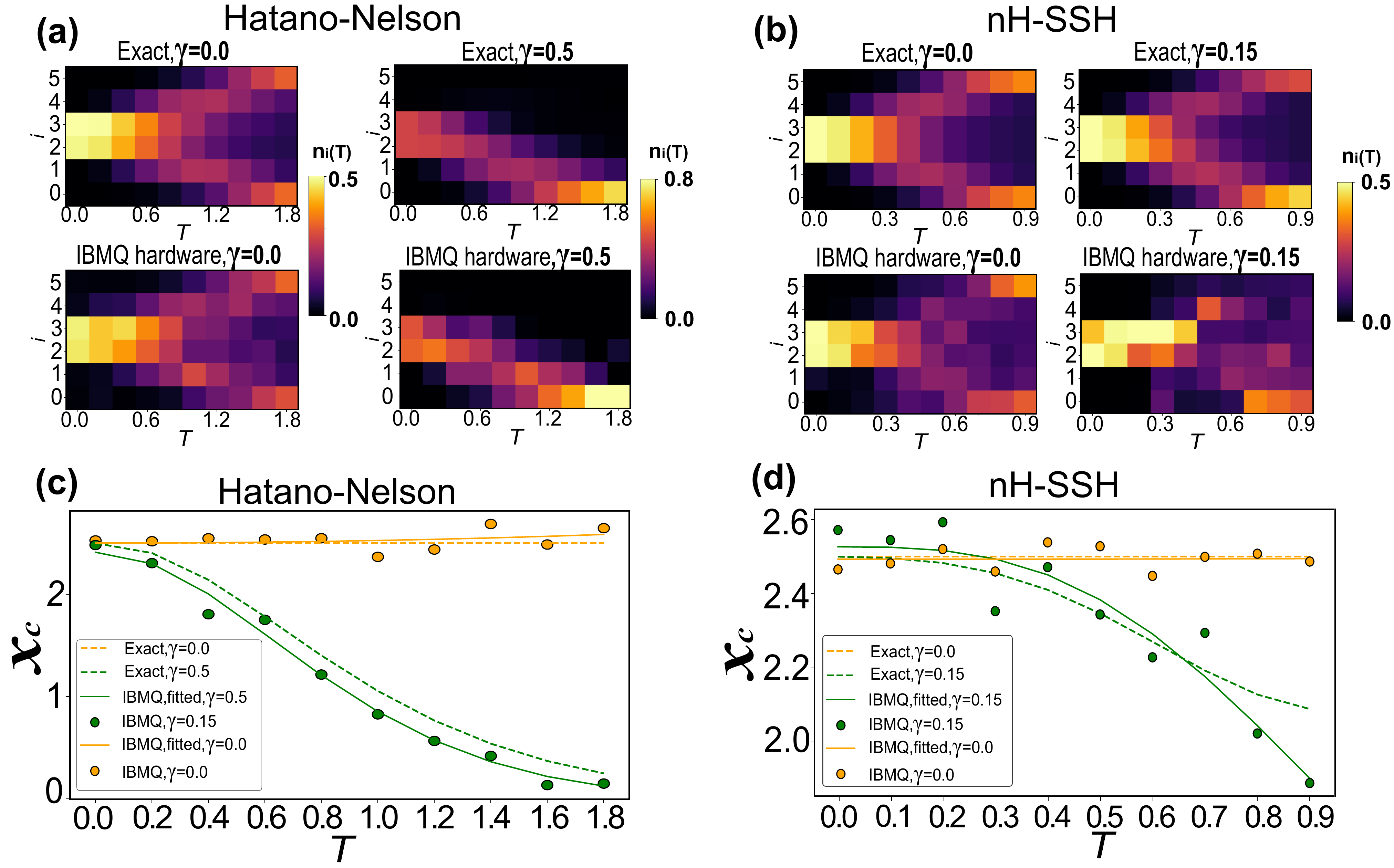}
	\caption{{\bf Classical and quantum simulations of single-particle non-Hermitian skin effect (NHSE) dynamics.} (a,b) The measured (IBM Q hardware) vs. classically simulated (Exact) time evolution of the fermion density $n_i(T)$, which is obtained from the qubit magnetization through Eq.~\ref{den}. 	
	Starting from the symmetrically-placed initial state $(\ket{\downarrow\downarrow\uparrow\downarrow\downarrow\downarrow}+\ket{\downarrow\downarrow\downarrow\uparrow\downarrow\downarrow})/\sqrt{2}$, the state evolves symmetrically in the absence of the NHSE ($\gamma=0.0$), but asymmetrically towards the $i=0$ boundary when the NHSE is switched on ($\gamma>0$). For the simpler Hatano-Nelson Hamiltonian (Eq.~\ref{spinssh_HN}), the asymmetric amplification is very evident with excellent agreement between simulation and measurements; for the more complicated non-Hermitian SSH Hamiltonian (Eq.~\ref{spinssh_main}), it is still visible amidst the noise.   
(c,d) Quantitative assessment of the asymmetric NHSE evolution through the density center-of-mass $x_C$ (Eq.~\ref{center}), based on the same data as in (a,b) above. In both models, negligible net propagation occurs without the NHSE (yellow);  in the presence of the NHSE (green), $x_c$ consistently evolves towards $0$, with very good agreement between exact diagonalization simulations and IBM Q measurements for the Hatano-Nelson model (c), and qualitative agreement for the nH-SSH model (d). These two models are respectively implemented on IBM superconducting circuits with 6 physical + 1 ancilla qubit, and 6 physical + 3 ancilla qubits, with $J=1.0$ and $2.0$ respectively. Each run is obtained over 160000 shots and conducted through variational quantum algorithms (see Methods). 
	}
	\label{fig:nhsshmain}
\end{figure*}

Next, to demonstrate the local decomposition approach (Eq.~\ref{hy2}), where each local non-Hermitian bond is assigned an ancilla qubit, we implement the prototypical non-Hermitian SSH chain~\cite{yao2018edge,lee2019anatomy}, which can be Jordan-Wigner transformed (see the Supplement) into the spin chain Hamiltonian 
\begin{equation}\label{spinssh_main}
	\begin{aligned}
		\hat{H}^{\rm Spin}_{\rm nH-SSH}=&-\sum_{j=0}^{L/2}\left((J+\gamma)\hat{X}^{+}_{2j}\hat{X}^{-}_{2j+1}+(J-\gamma)\hat{X}^{-}_{2j}\!\hat{X}^{+}_{2j+1}\right)\\
		&-\frac{J}{2}\sum_{j=0}^{L/2-1}\left(\hat{X}_{2j+1}\hat{X}_{2j+2}+\hat{Y}_{2j+1}\hat{Y}_{2j+2}\right),
	\end{aligned}
\end{equation} 
where $X_j=\ket{\uparrow}_j\bra{\downarrow}_j+\ket{\downarrow}_j\bra{\uparrow}_j$ and $Y_j=-i\ket{\uparrow}_j\bra{\downarrow}_j+\ket{\downarrow}_j\bra{\uparrow}_j$ are the $X$ and $Y$ Pauli matrix operators at site $j$. While this model is well-known to also exhibit topological boundary modes, its dominant dynamical behavior, which we are focusing on, is its asymmetric state pumping from the NHSE. To implement $\hat{H}^{\rm Spin}_{\rm nH-SSH}$ in a quantum circuit, note that its spin-spin couplings alternate between being Hermitian ($-J/2$) and non-Hermitian $-(J\pm \gamma)$, forming two-qubit unit cells, and thus admits the Trotter decomposition (Eq.~\ref{hy2}, sketched in FIG.~\ref{Hy_main}d) given by
\begin{equation}\label{dynamics}
	\begin{aligned}
		e^{-iT \hat{H}^{\rm Spin}_{\rm nH-SSH}}
		\approx[(\prod_{j}U^{\rm H}_{2j+1})(\prod_{j}U^{\rm nonH}_{2j})]^{\frac{T}{\delta t}},
	\end{aligned}
\end{equation} with the stacked unitary $U^{\rm H}_{j}=e^{+iJ\delta t( \hat{X}_{j+1}\!\hat{X}_{j}+\hat{Y}_{j+1}\!\hat{Y}_{j})}$ and non-unitary $U^{\rm nonH}_{j}=e^{+i\delta t( (J-\gamma)\hat{X}^{+}_{j+1}\!\hat{X}^{-}_{j}+(J+\gamma)\hat{X}^{-}_{j+1}\!\hat{X}^{+}_{j})}$. We insert an ancilla qubit in each unit cell so that its local two-qubit non-unitary operator $U^{\rm nonH}_{j}$ (blue shaded rectangle in FIG.~\ref{Hy_main}d) can be embedded within a three-qubit unitary operator (green shaded rectangle), as elaborated in the Methods.

To most vividly observe the dynamical consequences of the NHSE, we implement both models under open boundary conditions (OBCs), where the asymmetrically pumped state accumulates as ``skin'' boundary states instead of interfering with itself across the otherwise periodic boundary. For NISQ-era quantum processors, a universal challenge lies in the isometry decomposition of circuits using basis gates. Default decompositions provided by Qiskit's transpile function often result in exceedingly deep circuits, which is not ideal for robust noisy simulations. To overcome these obstacles, we employ variational quantum algorithms (VQAs) and optimize stacked ansatz circuits~\cite{koh2022stabilizing,chen2022high}, providing for an approximation of the target dynamics with shorter and more manageable circuit structure (See Methods). Since CX gates are a primary source of gate errors, our approach offers a significant advantage by substantially reducing the number of CX gates compared to the default decomposition provided by Qiskit's tool. Moreover, the reduced circuit depth in our optimized circuits helps to mitigate the impact of qubit decoherence.

\subsection{Measured dynamical NHSE behavior}

Using the above framework, we implemented the time evolution operator for the Hatano-Nelson and non-Hermitian SSH Hamiltonians on an IBM quantum processor, 
for its low error rates of single and controlled gates.

Below we report, for the first time in any digital quantum device, very prominent time-evolution signatures of the NHSE, observed for both the single- and multiple-fermion cases based on the measured pumping and accumulation of the fermion density along a qubit chain. At any measurement time $T$, the effective Jordan-Wigner transformed fermion density $n_i(T)$ at the $i$-th qubit (we index the qubits with $i$ instead of $x$ since they may not be linearly arranged in physical space) can be obtained from the measured magnetization $\hat{Z}_i=\ket{\uparrow}_i\bra{\uparrow}_i-\ket{\downarrow}_i\bra{\downarrow}_i$:
\begin{equation}\label{den}
	n_{i}(T)=|\bra{\psi(T)}\ket{\uparrow}_i|^2=\frac1{2}\left(\bra{\psi(T)} \hat{Z}_{i}\ket{\psi(T)}+1\right)
\end{equation}
where $\ket{\psi(T)} = e^{-i\hat HT}\ket{\psi(0)}$, and $\ket{\psi(0)}$ is the initial state prepared by appropriately flipping the qubits with $X$ gates. For each data point $n_i(T)$ in the subsequent plots, we execute our optimized time-evolution circuit a large number of times (160,000 per data point) and record the measurement strings. Post-selection is effected by discarding all strings in which the outcome at any ancilla qubit position is $\downarrow$.

\subsubsection{Single-fermion NHSE pumping}

We first attempt to observe the usual (single-particle) NHSE characterized by directional bulk currents by initializing the dynamics from the initially prepared state $\psi(0)=(\ket{\downarrow\downarrow\uparrow\downarrow\downarrow\downarrow}+\ket{\downarrow\downarrow\downarrow\uparrow\downarrow\downarrow})/\sqrt{2}$, which represents a single fermion (single $\ket\uparrow$) symmetrically superposed at the middle of a 6-site system indexed by $i=0,1,...,5$. FIGs.~\ref{fig:nhsshmain}a,b shows the time-evolution of the spatial density $n_i(T)$ subject to the Hatano-Nelson and non-Hermitian SSH Hamiltonians $\hat{H}^{\rm Spin}_{\rm HN}$ and $\hat{H}^{\rm Spin}_{\rm nH-SSH}$ respectively. In each case, the measured results (IBM Q hardware) are compared with their reference values from exact diagonalization (Exact), both when the non-Hermiticity is turned on ($\gamma \neq 0$) and off ($\gamma=0$).

For both models, by contrasting with the Hermitian $\gamma=0$ scenarios, it is evident that the NHSE leads to prominent state pumping towards the boundary $i=0$ site, accompanied by gain. This is corroborated by the reference results from exact diagonalization. In the Hatano-Nelson model, the initial state spreads out symmetrically when $\gamma=0$, but for $\gamma=0.5$, it is strongly pumped towards $i=0$ as time $T$ evolves, continuously amplified in the process. In the nH-SSH model, the directed pumping is not as clean due to the greater hopping complexity and interplay with topology; but even then, the final measured density still clearly turns out higher at the left boundary qubit $i=0$ than the right boundary qubit $i=5$, a testimony to the robustness of our approach in observing the NHSE.

To quantify the accuracy of our observed density pumping and to demonstrate that it goes far beyond statistical noise, we plotted in FIGs.~\ref{fig:nhsshmain}c,d the evolutions of the logical center of mass
\begin{equation}\label{center}
	x_{c}=\sum_{i=0}^{L-1}in_{i}.
\end{equation}
for both Hamiltonian evolutions, where $L=6$. As expected, under $\gamma=0$, there is no NHSE and $x_c$ (yellow) remains effectively constant for both models. As this is a case of unitary dynamics, no additional post-selection effort is required and the relatively small random errors arise from the inherent noise in the hardware. After the non-Hermiticity $\gamma$ is switched on, the center of mass $x_c$ (green) indeed curves towards boundary site $i=0$ for both models, significantly beyond the amplitude of the random deviations caused by noise.  In FIG.~\ref{fig:nhsshmain}c for the Hatano-Nelson model, the deviation of the measured $x_c$ (filled circles) from the exact reference trajectory (dashed) remains small, demonstrating that the NHSE can be accurately implemented and observed with our approach. In FIG.~\ref{fig:nhsshmain} (d) for the non-Hermitian SSH model, the errors are larger, but the qualitative directional pumping remains robust. The presence of larger noisy deviations is due to the utilization of one ancilla qubit per unit cell per time step in the ``local decomposition approach'', as compared to one ancilla qubit for the whole system per time step in the ``global approach'' for the Hatano-Nelson model. With more ancilla qubits come greater readout noise, but it is commendable that, despite the significant total amount of noise and decoherence in our NISQ-era quantum processor, we are ultimately able to compellingly observe the signature of NHSE pumping.

\begin{figure*}
	\centering
	\includegraphics[width=0.9\linewidth]{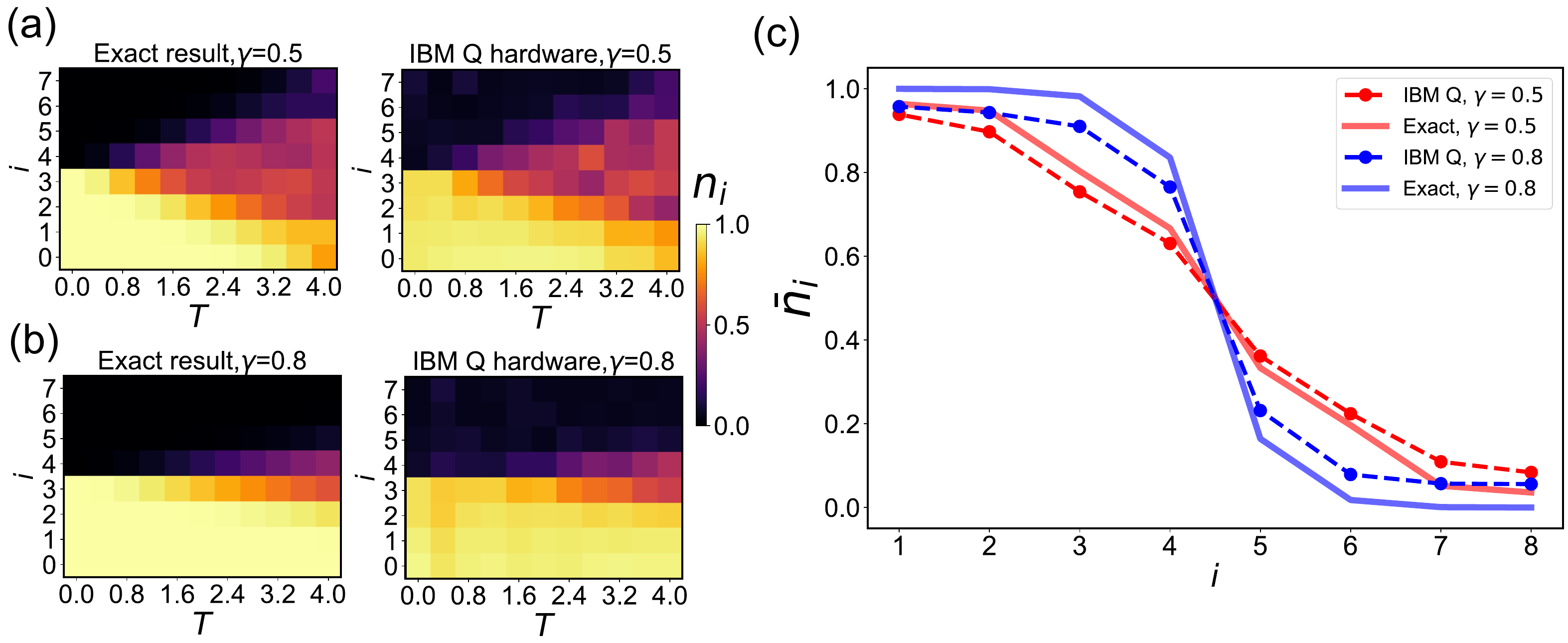}
	\caption{ \rz{{\bf Classical and quantum simulations of many-body NHSE dynamics and the emergence of the Fermi skin profile.}~(a,b) The good agreement between the measured (IBM Q hardware) vs. classically simulated (exact) time evolution of the fermion density $n_i(T)$ for our Hamiltonian Eq.~\ref{spinssh_main}, with the initial state being half-filling $\ket{\uparrow\uparrow\uparrow\uparrow\downarrow\downarrow\downarrow\downarrow}$ ($\ket{11110000}$ in the fermion basis). 
Under the weaker NHSE ($\gamma=0.5$), the initial state spreads out; but in the presence of the stronger NHSE ($\gamma=0.8$), the fermions are strongly localized near the edge qubit $i=0$. Notably, with multiple fermions, Pauli exclusion prevents complete collapse towards $i=0$, instead giving rise to an eventual profile $n_i(T)$ that transitions from almost maximal ($n_i\approx 1$) to almost vanishing ($n_i\approx 0$) as $i$ increases, reminiscent of the Fermi-Dirac distribution in real space (Eq.~\ref{nFs}). 
(c) This ``Fermi skin'' profile for the half-filling case with NHSE is more clearly observable through the time-averaged density $\bar{n}_{i}$ (Eq.~\ref{tavm}) under $\gamma=0.5$ (red) and $\gamma=0.8$ (blue), shown here averaged over time intervals $T_{\rm total}=4.0$, with good agreement between IBM Q measurements (dashed curve) and exact diagonalization (solid curve) simulations. For all results, we set $J=1.0$ and $\delta t=0.1$. IBM Q measurements are averaged over 160000 shots for each Trotter step.}
}
	\label{fig:111000}
\end{figure*}

\subsubsection{Fermi skin from multiple-fermion NHSE}
Having conclusively observed the NHSE of a single fermion, we next observe the emergence of the Fermi skin density profile, a novel many-body signature of the NHSE that exists only in the presence of multiple fermions. If each single-fermion eigenstate is exponentially localized by the NHSE, it can be shown by explicit evaluation of their Slater determinant (see Methods) that, after asymptotically long times, the real-space many-body density profile $n_\text{Fs}(i)$ takes the approximate Fermi-Dirac form 
\begin{equation}
n_\text{Fs}(i) \approx \frac1{1+e^{\beta_\text{eff}(i-N)}}, \qquad i=0,...,L-1
\label{nFs}
\end{equation}
where the ``chemical potential'' $N$ corresponds to the number of fermions and $\beta_\text{eff}\propto\kappa=\log(\frac{J+\gamma}{J-\gamma})$ is its effective inverse ``temperature'' that scales with the NHSE hopping asymmetry. \rz{While this Fermi-Dirac approximation ansatz becomes strictly accurate at larger $\kappa$, its agreement with simulation data is already on par with the noise tolerance from the quantum hardware for our values of $\kappa$ used.}
This real-space Fermi-Dirac profile results from the competition between Fermi degeneracy pressure (Pauli exclusion) and the NHSE, in analogy to the Fermi surface of a thermal Fermi gas, which is borne out of the competition between Fermi degeneracy pressure and the Boltzmann weight. In the NHSE, each solitary fermion occupies the site $i$ with a probability $e^{-\alpha i}$; in thermodynamics, it occupies the energy level $E_i$ with the Boltzmann weight $e^{-\beta E_i}$.

On a quantum processor, demonstrating this many-body Fermi skin would require little additional effort beyond realizing the single-particle NHSE: one just has to prepare an initial state $\psi(0)$ with multiple up-spins, and confine final measurements to this $\sum_i\hat Z_i$-conserved sector. \rz{Yet, observing the Fermi skin would have been impossible in existing experiments that demonstrated the NHSE through classical platforms~\cite{zou2021observation}.}

\rz{To capture the signatures of such an edge Fermi skin on a quantum processor, we simulate the time evolution of a half-filled initial state, $\ket{\psi(0)}=\ket{\uparrow\uparrow\uparrow\uparrow\downarrow\downarrow\downarrow\downarrow}$, as shown in FIGs.~\ref{fig:111000} (a) and (b), where the quantum hardware simulations agree excellently with reference results from exact diagonalization. Under relatively weaker NHSE ($\gamma/J=0.5$ for FIG.~\ref{fig:111000} (a)), fermions initially stabilize near the boundary but gradually spread out over time. However, when the NHSE strength is increased to $\gamma/J=0.8$, all fermions remain strongly localized for the entire simulation duration, as shown in FIG.~\ref{fig:111000} (b). This presents a stark contrast to the usual NHSE in classical or single-particle settings, which usually results in an exponentially localized boundary ``skin" profile.}

\rz{To check whether our observed fermion density profile is indeed of Fermi skin form, we even out the temporal fluctuations by considering the time-averaged spatial fermion density [FIG.~\ref{fig:111000} (c)]
\begin{equation}\label{tavm}
	\bar{n}_{i}=\frac1{N_{\rm steps}}\sum^{N_{\rm  steps}}_{N=1}n_{i}(N\delta t),
\end{equation}
which averages the many-body density $n_i$ over all the time steps of $\delta t= 0.1$ interval, where $N_{\rm steps}=T_{\rm total}/\delta t$ is the total number of Trotter steps and $T_{\rm total}=4.0$ is the total duration of the state evolution used.  Here, under $\gamma/J=0.8$, the Fermi skin (blue curves), i.e., the symmetric Fermi-Dirac-like profile, can still be clearly observed.}

\rz{\subsubsection{Bulk Fermi skin from repulsive interactions}
\begin{figure*}
	\centering
	\includegraphics[width=0.8\linewidth]{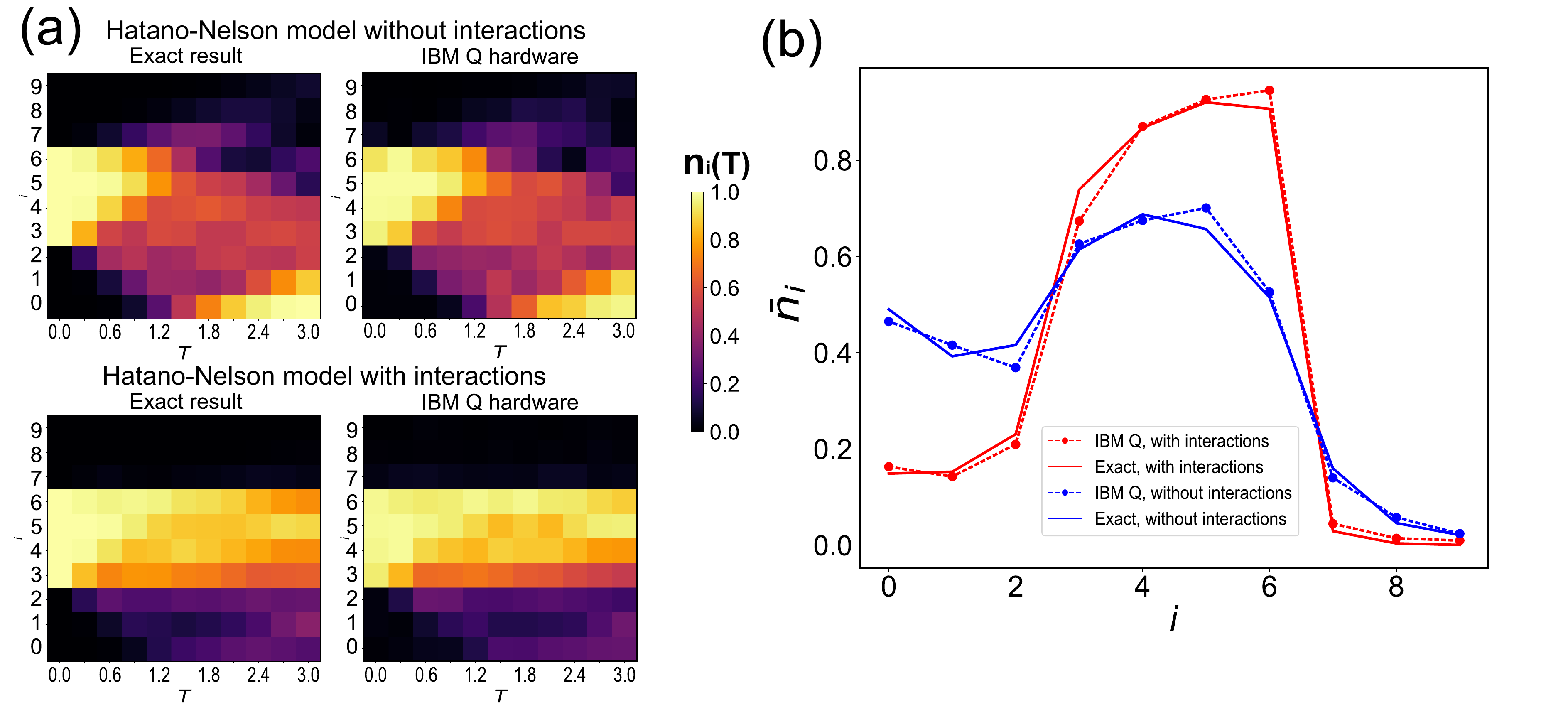}
	\caption{\rz{{\bf Quantum simulations of many-fermion NHSE dynamics and the interaction-induced bulk Fermi skin.} We simulate the time evolution of the fermion density $n_i(T)$ (color bar shown in panel (a)) of the interacting and non-interacting models described by Eq.~\ref{int} and Eq.~\ref{spinssh_HN} respectively. The initial state is $\ket{\downarrow\downarrow\uparrow\uparrow\uparrow\uparrow\downarrow\downarrow}$. Panel (a) shows the comparison between experimental measurements on IBM Q hardware and classically simulated (exact) results, with excellent agreement. The strong interactions of $U/J=5.0$ significantly suppress the asymmetric propagation, stabilizing the Fermi skin at the center. Panel (b) provides a quantitative assessment of the dynamical suppression shown in panel (a) through the time-averaged density $\bar{n}_{i}$ (Eq.~\ref{tavm}). Notably, the comparison between exact (solid curve) and simulated (dots) results demonstrates a strong agreement. The red curve in particular highlights the bulk Fermi skin with slight asymmetric propagation.
IBM Q simulations are implemented on circuits with $8$ physical + $1$ ancilla qubits. We set $J=1.0$, $\gamma=0.5$, and $U=5.0$.
Each data point is averaged over $160000$ shots, to minimize sampling errors.}
}
	\label{fig:dynamics3}
\end{figure*}
Previously, we proposed the emergence of bulk Fermi skin with fermions that become clustered due to interactions. This phenomenon can exhibit reversed skin localization compared to edge Fermi skin due to kinetic constraints. To achieve such particle clustering, we introduce an additional nearest-neighbor interaction between the nearest $\ket{\uparrow}$ levels of qubits:
\begin{equation}\label{int}
\hat{H}_{\rm int}=\hat{H}^{\rm Spin}_{\rm HN}+U\sum^{L-1}_{i}\hat{n}_{i} \hat{n}_{i+1}
\end{equation}
with $\hat{n}_i=\ket{\uparrow}_i\bra{\uparrow}_i$. To simulate this model, we modify the unitary operator in FIG.~\ref{Hy_main} to $U^{\rm nonH \prime}_{j}=e^{+i\delta t\left[ (J-\gamma)\hat{X}^{+}_{j+1}\!\hat{X}^{-}_{j}+(J+\gamma)\hat{X}^{-}_{j+1}\!\hat{X}^{+}_{j}-U\hat{n}_{j} \hat{n}_{j+1}\right]}$. The implementation of this model on the IBM Q device follows the procedure detailed for FIG.~\ref{fig:111000}. 
}

\rz{In our simulatiosn, we consider a strong interaction of $U/J=5$. Such interactions can naturally induce an energy gap between configurations with clustered and spatially separated fermions. Accordingly, we initialize the system in the state $\ket{\downarrow\downarrow\uparrow\uparrow\uparrow\uparrow\downarrow\downarrow}$, representing nearest-neighbor fermions at the center of the lattice. This particular initial state significantly overlaps with the configuration with extremal energy. As a result, the interaction-induced gap can lead to the suppression of overall propagation. This behavior is demonstrated in FIG.~\ref{fig:dynamics3}a, where strong interactions notably stabilize fermions in the bulk. Such density extends towards the amplification direction, induced by ($t+\gamma$) terms.
We also average the many-body density $n_i$ over $11$ steps in the duration $T_{\rm total}=3.0$, as shown in FIG.~\ref{fig:dynamics3}b. Notably, the red curve for the interacting model depicts our proposed bulk Fermi skin (see FIG.~\ref{fig:nhsemain}c), where skin localization emerges in the bulk and extends towards the amplification of non-reciprocal couplings.}

%=====================================================Conclusions===================================================================
\subsection{Conclusions}

In this work, we report the first demonstration of the NHSE on a NISQ-era quantum computer, through both conventional single-particle dynamics and its elusive many-fermion Fermi skin profile. Moreover, our quantum circuits with embedded non-unitary components provide a promising scalable approach toward the quantum simulation of non-Hermitian Hamiltonians. Thanks to the intrinsic quantum nature of digital quantum computers, our approach is particularly suitable for the simulations of new non-Hermitian phenomena involving many-body statistics and interactions, an uncharted terrain in experimental investigations. Furthermore, our high-fidelity simulations through the deployment of VQAs provide valuable insights into the great potential of variational optimization techniques in quantum simulation.

Even though this present study primarily focuses on the specific demonstration of the NHSE and its interplay with many-body statistics, with the remarkable programmability of quantum computers, our approach can be universally adapted to physically demonstrate a variety of other novel non-Hermitian dynamical phenomena such as non-unitary criticality, non-Hermitian skin clustering and non-Hermitian many-body localization~\cite{PhysRevLett.123.090603,mak2024statics,shen2022non} .
\\

{\bf Author contributions.} R.~S  and T.~C  contributed equally to this work. C.~H.~L proposed the initial idea, developed the Fermi skin formalism, guided the overall research direction, and supervised this work. R.~S and T.~C developed the algorithms and implemented them on the IBM Q devices together. Y. B contributed to numerical algorithms. All the authors contributed to the manuscript.

{\bf Data availability.}
All data of this work are available from the corresponding authors upon reasonable request.

{\bf Code availability.}
All codes of results in this work are available from the corresponding authors upon reasonable request.
\begin{acknowledgements}
	{\bf Acknowledgements.}	 T.~C. and R.~S. thank Truman Ng for discussions on the quantum simulation implementation on IBM Quantum services. C.~H.~L. and T.~C. acknowledge support from Singapore's NRF Quantum engineering grant NRF2021-QEP2-02-P09 and Singapore's MOE Tier-II grant Proposal ID: T2EP50222-0008. T.~C. and B.~Y. acknowledges the support from the Singapore National Research Foundation (NRF) under the NRF fellowship award NRF-NRFF12-2020-0005. We acknowledge the use of IBM Quantum services for this work. The views expressed are those of the authors and do not reflect the official policy or position of IBM or the IBM Quantum team.
\end{acknowledgements}

%================================================Method==============================================================================
\section{Methods}

\subsection{Implementing nonunitary processes in quantum circuits} 
As discussed in the main text, since the native gates in existing quantum processors are unitary, non-unitary operators cannot be directly implemented in quantum circuits by the conventional method of operation decomposition. In our work, we employ post-selection to realize the non-unitary dynamics of our Hamiltonians, where an $N$-qubit non-unitary operation can be embedded in an $(N+1)$-qubit unitary operator. We first provide a single-qubit example for an illustrative Hamiltonian given by $\hat H_\pm = \frac{i}{2}(I_{2\times 2}\mp \hat Z)$. Under the evolution of a parameter $\phi$, i.e. time, we have the non-unitary evolution operators
\begin{equation}
	R_{+}=\left[\begin{array}{cc}
		1 & 0 \\
		0 & e^{-\phi}
	\end{array}\right],\qquad 	R_{-}=\left[\begin{array}{cc}
		e^{-\phi} & 0 \\
		0 & 1
	\end{array}\right].
\end{equation}
%imaginary fields $\frac{\pm\phi i}{2} \hat{Z}$, which leads to the non-unitary evolution $e^{\mp\frac{\phi}{2} \hat{Z}}$. Such single-qubit non-unitary operators can be embedded in two-qubit unitary operators $U_{R^{\pm}}$ by the following ansatz $U_{R_{\pm}}$:
The key idea in the post-selection approach is to embed this non-unitary $R_\pm$ into a larger unitary matrix $U_{R_\pm}$, such that the non-unitary evolution can be extracted from the larger unitary evolution operator via post-selection. Considering only one ancilla qubit for simplicity, $U_{R_\pm}$ is doubled in size compared to $R_\pm$, and in this case, with a single physical qubit, takes the $4\times 4$ form
\begin{equation}
	U_{R_{\pm}}=\left[\begin{array}{cc}
		R_{\pm} & B_{2\times2} \\
		C_{2\times2} & D_{2\times2}
	\end{array}\right],
\end{equation}
where submatrices $ B_{2\times2},C_{2\times2}$ and $D_{2\times2}$ are to be determined. The unitarity condition $U^{\dagger}_{R^{\pm}}U_{R^{\pm}}=I_{\rm 4\times4}$ leads to the constraint $C_{2\times2}C^{\dagger}_{2\times2}+R_{\pm}R^{\dagger}_{\pm}=I_{\rm 2\times2}$. This allows for an explicit construction of an admissible $C_{2\times 2}$ such as $C_{2\times2}=-  \sqrt{I_{\rm 2\times2}-R_{\pm}R^{\dagger}_{\pm}}$, which can be analytically computed.

By setting $B_{2\times2}=-C_{2\times2}$ and $D_{2\times2}=R_{\pm}$, we obtain
\begin{equation}\label{lossmain}
	U_{R_{-}}=\left[
	\begin{array}{cccc}
		e^{-\phi} & 0 & \sqrt{1-e^{-2\phi}} & 0 \\
		0 & 1 & 0 & 0 \\
		-\sqrt{1-e^{-2\phi}} & 0 & e^{-\phi} & 0 \\
		0 & 0 & 0 & 1
	\end{array}
	\right],
\end{equation}
and
\begin{equation}\label{gainmain}
	U_{R_{+}}=\left[
	\begin{array}{cccc}
		1 & 0 & 0 & 0 \\
		0 & e^{-\phi} & 0 & \sqrt{1-e^{-2\phi}} \\
		0 & 0 & 1 & 0 \\
		0 & -\sqrt{1-e^{-2\phi}} & 0 & e^{-\phi}
	\end{array}
	\right]
\end{equation} 
which is unitary as constructed.

In the following, we illustrate how to post-select the ancilla qubits to isolate the non-unitary evolution $R_\pm$. We follow the Qiskit convention where higher qubit indices are more significant (little-endian convention)~\cite{Qiskit}. We prepare a the following two-qubit state to act on $U_{R^{-}}$ in Eq.~\ref{lossmain}:
\begin{equation}
	% \ket{\psi}\otimes\ket{0}_{A}=\begin{pmatrix}
		% 	a&0&b&0
		% \end{pmatrix}^{T},
	\ket{\uparrow}_{A^{-}}\otimes\ket{\psi}=\begin{pmatrix}\ket{\psi},&0\end{pmatrix}^{T},
\end{equation}
where $\ket{\psi}$ is the physical single-qubit state, and ${\ket{0}_{A^{-}}}$ represents the ancilla state. Similarly, for $U_{R^{+}}$ in Eq.~\ref{gainmain}, we prepare
\begin{equation}
	% \ket{\psi}\otimes\ket{0}_{A}=\begin{pmatrix}
		% 	a&0&b&0
		% \end{pmatrix}^{T},
	\ket{\downarrow}_{A^{+}}\otimes\ket{\psi}=(0,\ket{\psi})^{T},
\end{equation}
but with the spin-down ancilla state $\ket{\downarrow}_{A^{+}}$. After applying $U_{R_{\pm}}$ on the above states, we obtain:
\begin{align}
	&U_{R^{-}}\left(\ket{\uparrow}_{A^{-}}\otimes\ket{\psi}\right)\rightarrow\ket{\uparrow}_{A^{-}} \otimes R_{-}\ket{\psi},\\ \nonumber
	&U_{R^{+}}\left(\ket{\downarrow}_{A^{+}}\otimes\ket{\psi}\right)\rightarrow\ket{\downarrow}_{A^{+}} \otimes R_{+}\ket{\psi},
\end{align}
from which the $R_\mp$ non-unitaries can be recovered by simply projecting out the $\ket{\uparrow}$ or $\ket{\downarrow}$spins in the ancilla qubit.

\subsection{Realizing generic non-unitary operators through post-selection} In the above, we provided an illustrative example of how to embed single-qubit non-unitary operators in a quantum circuit. Analogously, to embed a desired two-qubit operator $U^{\rm nonH}$ corresponding to a non-Hermitian term into a unitary three-qubit unitary operator $U_{R^{\prime\prime}}$, we can use the generalized ansatz
\begin{equation}\label{rpp}
	U_{R^{\prime\prime}}=\left[\begin{array}{cc}
		uU^{\rm nonH} & B_{4\times4} \\
				C_{4\times4} & D_{4\times4}
	\end{array}\right],
\end{equation}
where $u^{-2}$ denotes the maximum eigenvalue of $U^{\rm nonH\dagger}U^{\rm nonH}$, and 
$C_{4\times4}=A\sqrt{I_{4\times4}-u^{2}\Sigma^{2}}B^{\dagger}$, where the singular value decomposition of $U^{\rm nonH}$ gives: $U^{\rm nonH}=A\Sigma B^{\dagger}$ ~\cite{lin2021real}.

Here, the rescaling factor $u$ is necessary for ensuring that $(I_{4\times4}-u^{2}\Sigma^{2})$ is real and non-negative. We can similarly solve for $B_{4\times4}$ and $D_{4\times4}$ numerically by QR decomposition of the following matrix
\begin{equation}\label{nonunssh}
	W_{8\times 8}^{\prime}=\left[\begin{array}{cc}
		uU^{\rm nonH} & I_{4\times4} \\
		C_{4\times4} & I_{4\times4}
	\end{array}\right]=U_{R^{\prime\prime}} M_{8\times8},
\end{equation}
where $M_{8\times8}$ is the upper triangular matrix from the QR decomposition of $W_{8\times 8}^{\prime}$.

To illustrate how such a three-qubit post-selection operation works in the quantum circuit, we consider a two-qubit subspace with an ancilla qubit expressed as $\ket{\psi^{\prime\prime}}\otimes{\ket{\uparrow}_{A}}$ (where $\ket{\psi^{\prime\prime}}$ is the physical state and $\ket{\uparrow}_{A}$ is the ancilla state), for which the above three-qubit operation in Eq.~\ref{rpp} leads to 
\begin{equation}\label{ancilla}
	U_{R^{\prime\prime}}\ket{\psi^{\prime\prime}}\otimes{\ket{\uparrow}_{A}}=(uU^{\rm nonH}\ket{\psi^{\prime\prime}})\otimes{\ket{\uparrow}_{A}}+(C\ket{\psi^{\prime\prime}})\otimes\ket{\downarrow}_{A},
\end{equation}
which recovers the action of $U^{\rm nonH}$ upon post-selection on $\ket{\uparrow}_{A}$. 

\begin{figure}
	\centering
	\includegraphics[width=0.99\linewidth]{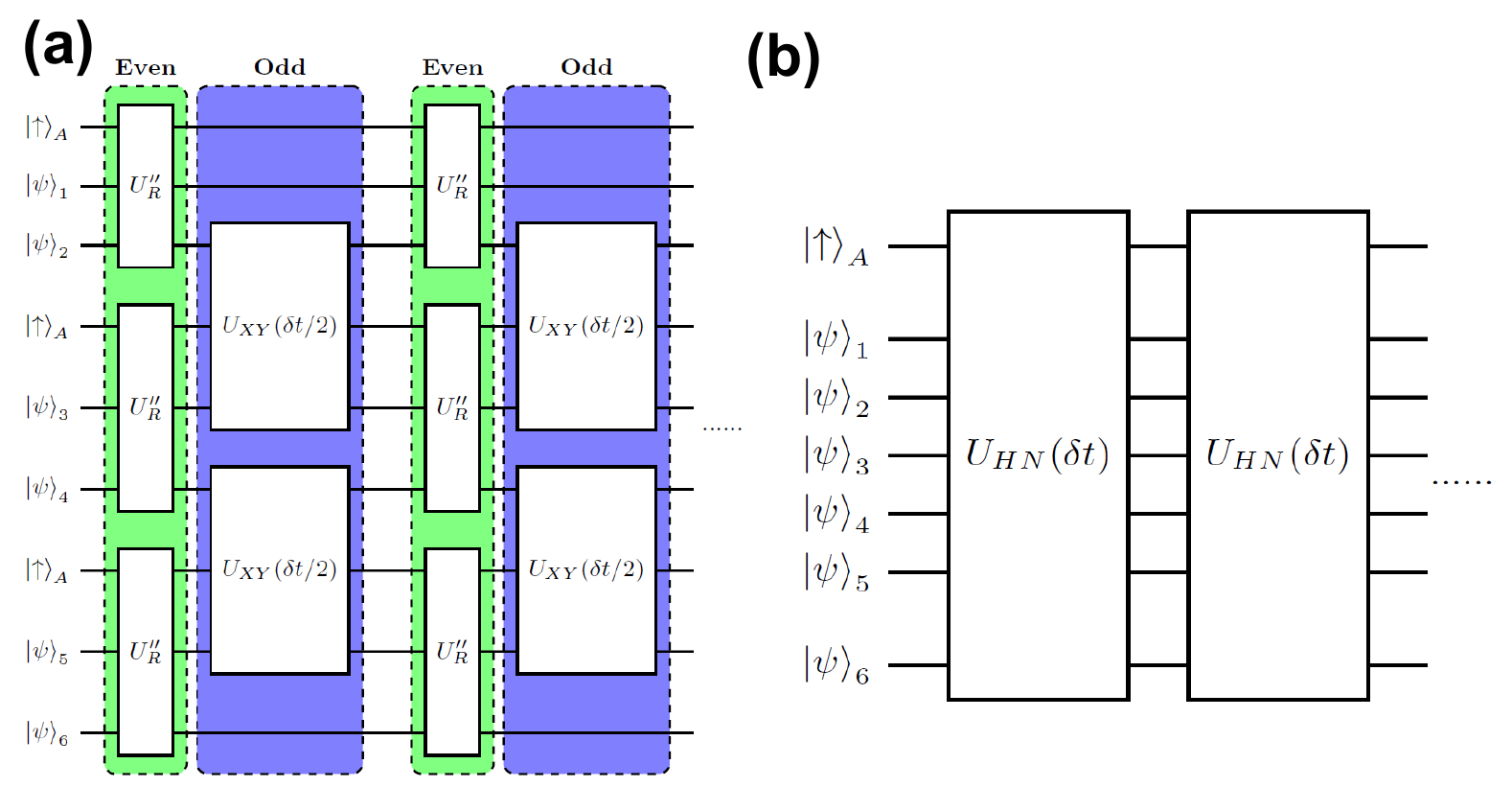}
	\caption{{\bf Structure of the quantum circuits for nH-SSH and Hatano-Nelson Hamiltonian evolutions.}~(a) The Trotterization structure of the nH-SSH model's quantum circuit (6 physical and 3 ancilla qubits), also shown schematically in FIG.~\ref{Hy_main}b,d of the main text. %The index $i$ for the physical qubit encoded as $\ket{\psi}_{i}$ represents its real-space position, and $\ket{\uparrow}_{A}$ is for the initial ancilla qubit indicated in Eq.~\ref{ancilla}.  
The subscripts $i=0,...,5$ label the qubits on the physical non-Hermitian SSH chain, while the subscript ``A'' labels the ancilla qubits. $U_{R^{\prime\prime}}$ (green) denotes the operation in Eq.~\ref{rpp}, which implements the 2-qubit non-unitary evolution. $U_{XY}$ (blue) represents the operation $e^{i\delta t\left(\frac{J}{2}\hat{X}\!\otimes\hat{I}\otimes\!\hat{X}\!+\!\frac{J}{2}\hat{Y}\!\otimes\hat{I}\otimes\!\hat{Y}\!\right)}$ that couples neighboring unit cells. (b) The structure of the Hatano Nelson model's quantum circuit (6 physical and 1 ancilla qubits), also shown schematically in FIG.~\ref{Hy_main}a,c in the main text. Each non-unitary Trotter step is realized by the ancilla-based implementation in Eq.~\ref{MHN}. For both models, two Trotter steps are shown.
	}
	\label{nhssh2}
\end{figure}

To realize the non-unitary dynamics $e^{-i\delta tH^{\rm spin}_{\rm nH-SSH}}$ of our non-Hermitian SSH model, we shown in FIG.~\ref{nhssh2} the circuit for 2 Trotter steps, each which involves 6 physical qubits and 3 ancilla qubits. %Note that the above operator $U_{R^{\prime\prime}}$ requires addressing an ancilla qubit as the first qubit of each three-qubit subspace, so for the left circuit in FIG.~\ref{recom}, we can insert an ancilla qubit in front of each physical two-qubit cell. 
%CH: This indexing is conceptually trivial
Each Trotter step is locally broken into even (non-unitary, green) and odd (unitary, blue) bonds as in Eq.~\ref{hy2} in the main text
\begin{equation}\label{trotterdy}
	\begin{aligned}
		&e^{+i\delta t\hat{H}^{\rm spin}_{\rm nH-SSH}}\approx [\prod_{j}e^{i\delta t(+\frac{J}{2}(\hat{X}_{2j+1}\hat{X}_{2j+2}+\hat{Y}_{2j+1}\hat{Y}_{2j+2}))}]\\ &[\prod_{j}e^{+i\delta t((J+\gamma)\hat{X}^{+}_{2j}\hat{X}^{-}_{2j+1}+(J-\gamma)\hat{X}^{-}_{2j}\!\!\hat{X}^{+}_{2j+1})}]=U_\text{odd}U_\text{even},
	\end{aligned}
\end{equation}
where $j$ is for the real-space position of the physical qubit $Q_{j}$, and $\hat{X}^{\pm}=\frac{\hat{X}\pm i\hat{Y}}{2}$. The non-unitary bonds in Eq.~\ref{trotterdy} are realized through Eq.~\ref{rpp} as described earlier.

To simulate the nonunitary dynamics of the Hatano-Nelson model, we globally attached 1 ancilla qubits to all 6 physical qubits, such that the entire Trotter step is embedded in the unitary
\begin{equation}\label{MHN}
	U_{\rm HN}=\left[\begin{array}{cc}
		uR_{\rm HN} & B_{64\times64} \\                          
		C_{64\times64} & D_{64\times64}
	\end{array}\right],
\end{equation}

%\begin{equation}\label{MHN}
%	U_{\rm HN}=\left[\begin{array}{cc}
%\textcolor{red}{e^{-i{\hat H}^\text{Spin}_\text{HN}\delta t}} & B \\                          
%		C & D
%	\end{array}\right],
%\end{equation}

which can be solved analogously as before, with $R_\text{HN}$ sketched in the main text and further elaborated in the Supplementary Materials.

%===============================================FIG 6==========================================================================
\begin{figure*}
	\centering
	\includegraphics[width=0.96\linewidth]{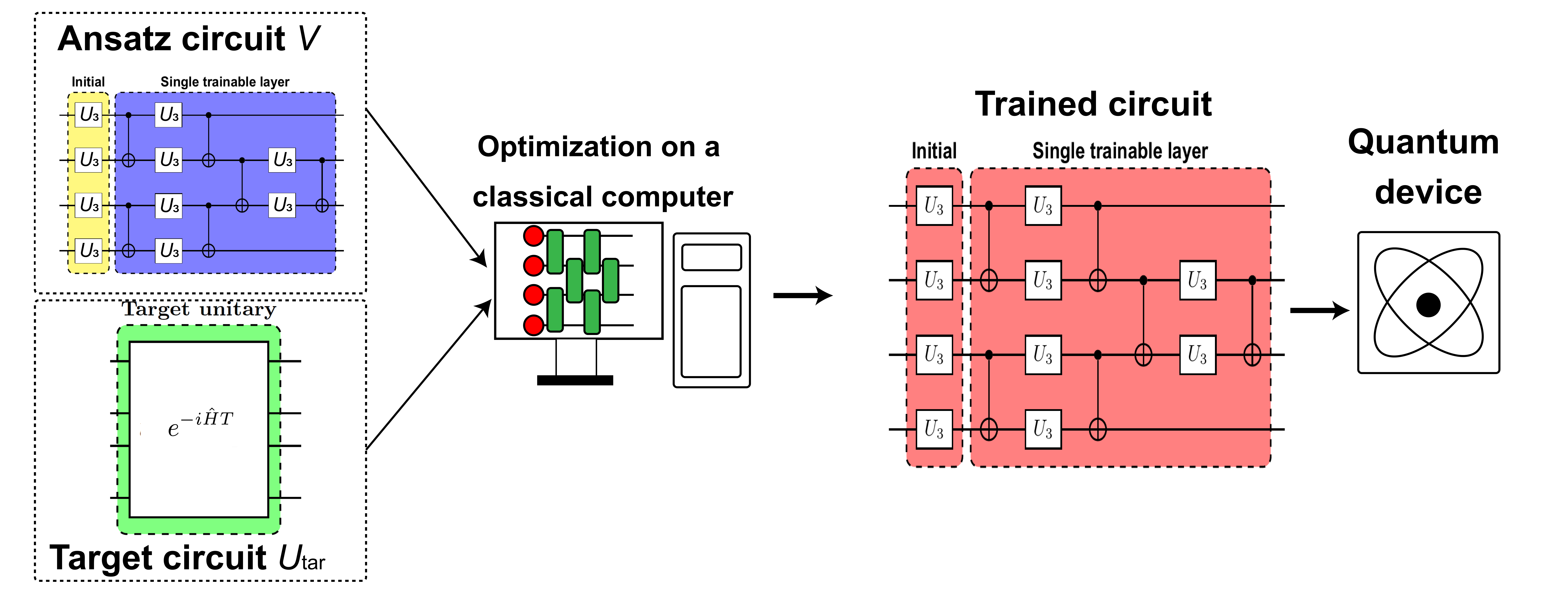}
		\caption{{\bf Workflow for our variational approach to non-unitary dynamics.} First, the target unitary $U_\text{tar}$, which effectively implements the process $e^{-i\hat HT}$, (green block) and the trainable unitary ansatz $V$ (blue block) are represented in a matrix product state (MPS) tensor network on a classical computer. The trainable unitary consists of several layers, and each layer consists of a fixed configuration of $U{3}$ and $CX$ gates. Next, the $U_{3}$ gate parameters in the trainable unitary are optimized by maximizing the overlap of the target and trainable unitaries with respect to the given initial state i.e. by minimizing the cost function in Eq.~\ref{Q}. The trained ansatz circuit is then transcribed onto the quantum circuit device.
		}
		\label{recom}
	\end{figure*}

\subsection{Variational quantum algorithms (VQAs) for noise mitigation}
To realize robust simulations on current noisy quantum devices, we use variational quantum algorithms (VQAs) to reduce circuit depth. We use a trainable ansatz circuit $V$ to approximate a desired target unitary circuit $U_{\rm tar}$. In this work, the target unitary 
$U_{\rm tar}$, which effectively implements the process $e^{-i\hat HT}$, is the complete time evolution circuit that acts on both the physical and ancilla qubits overall $N_\text{steps}=T/\delta t$ Trotter steps, as shown in FIGs.~\ref{Hy_main} and \ref{nhssh2} for our two illustrative models.  For the non-Hermitian SSH model, the target $U_{\rm tar}$ is built according to
\begin{equation}
	U^\text{nH-SSH}_{\rm tar}(T)=(U_{\rm odd}U_{\rm even})^{N_\text{steps}}, 
\end{equation}
where $U_{\rm even}$ and $U_{\rm odd}$ denote the even (green) blocks and odd (blue) blocks shown in FIG.~\ref{nhssh2}a. The target unitary for our illustrative Hatano-Nelson model 
\begin{equation}
U^\text{HN}_{\rm tar}(T)=U^{N_\text{steps}}_{\rm HN}
\end{equation} 
is simply the single Trotter step $U_{\rm HN}$ (Eq.~\ref{MHN}) raised to the $N_\text{steps}$-th power (FIG.~\ref{nhssh2}b). These are further detailed in the Supplementary Materials.

\begin{figure}[h]
	\centering
	\includegraphics[width=\linewidth]{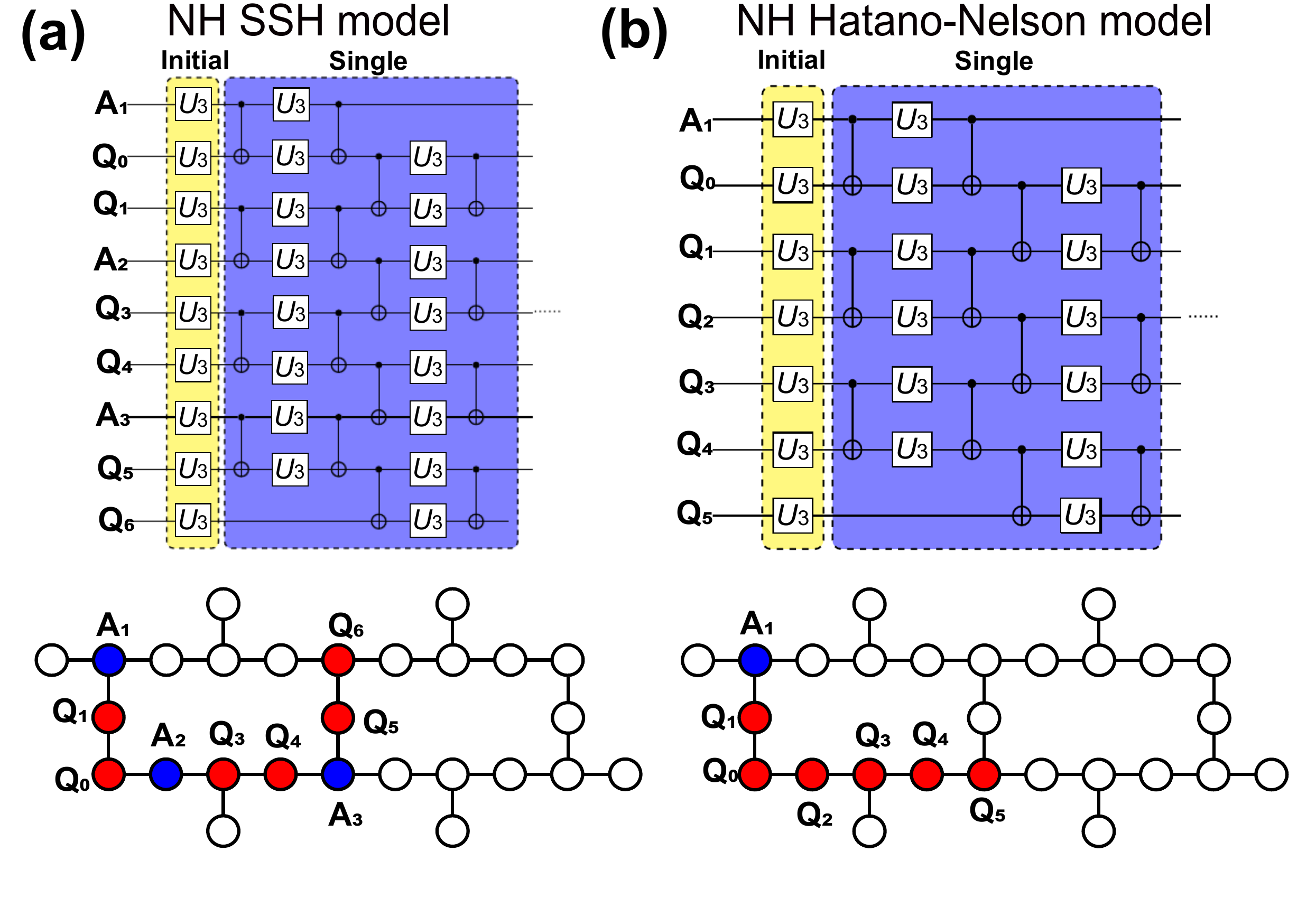}
	\caption{{\bf Implementation of optimized quantum circuit ansatz.} The trainable ansatz circuits $V$ for our (a) non-Hermitian SSH chain and (b) non-Hermitian Hatano-Nelson chain each consists of an initial $U_3$ gate and multiple ansatz layers that can be optimized. While each model requires different numbers of ancilla ``A'' qubits, their ansatz circuits share the same structure, apart from their overall size. Their physical transpilations onto the IBM quantum processor ``ibmq\_toronto" are shown at the bottom.
	} 
	\label{fig:devicemaina}
\end{figure}

To greatly reduce noise, an ansatz circuit $V$ of fixed depth and structure is trained such as to optimally approximate $U_{\rm tar}$, and then fed into the noisy quantum device. The training workflow is schematically illustrated in FIG.~\ref{recom}. Structure-wise, as depicted in FIG.~\ref{fig:devicemaina}, our trainable circuit consists of multiple ansatz layers (blue), each consisting of an alternating arrangement of $U_3$ single-qubit rotations and two-qubit $CX=I \otimes|\uparrow\rangle\langle\uparrow|+X \otimes|\downarrow\rangle\langle\downarrow|$ gates between neighboring qubits. As shown in FIG.~\ref{fig:devicemaina}a for our illustrative non-Hermitian SSH evolution operator with $L=6$ physical qubits and $L/2=3$ ancilla qubits, there are $(3L-2)=16$ $CX$ gates per layer.

Since the structure of the ansatz circuit $V$ is fixed, its trainable parts are its $U_{3}$ gates given by~\cite{Qiskit}:
\begin{equation}
	U_{3}(\theta, \phi, \lambda)=\left[\begin{array}{cc}
		\cos \left(\frac{\theta}{2}\right) & -e^{i \lambda} \sin \left(\frac{\theta}{2}\right) \\
		e^{i \phi} \sin \left(\frac{\theta}{2}\right) & e^{i(\phi+\lambda)} \cos \left(\frac{\theta}{2}\right)
	\end{array}\right].
\end{equation}
Each $U_3$ gate contains three angular parameters to be optimized, and the set of parameters $({\bm \theta, \bm\phi, \bm\lambda})$ across all the $U_{3}$ gates are obtained by minimizing the following cost function
\begin{equation}\label{Q}
	Q({\bm \theta, \bm\phi, \bm\lambda})=1-|\bra{\psi_{0}}V^{\dagger}(\theta, \phi, \lambda)U_{\bf tar}\ket{\psi_{0}}|,
\end{equation}
where $\ket{\psi_{0}}$ is the initial state. Here, the operator $V^{\dagger}({\bm \theta, \bm\phi, \bm\lambda})U_{\rm tar}$ is expressed as a matrix product operator in the cost function, which is minimized when $V(\theta, \phi, \lambda)\ket{\psi_{0}}$ and $U_{\bf tar}\ket{\psi_{0}}$ overlap as much as possible. Note that in the matrix product operator for $U_{\rm tar}$, an additional projection operator $\ket{\uparrow}\bra{\uparrow}$ needs to be applied on each ancilla qubit at each Trotter step. Our VQA circuits require fewer $CX$ gates than the general Trotterization approach and help make it possible to achieve good signal-to-noise ratios amidst significant device noise.

\rz{In our work, we consider the systems sizes up to $13$ qubits. At this scale, it is crucial to maintain a balance between the effect of noise and circuit depth to observe meaningful results. To this end, we set the circuit depth to include up to 8 ansatz layers (illustrated as the blue block shown in FIG.~\ref{fig:devicemaina}). This setup ensures that the cost function can attain a convergence of more than $90\%$. As such, this optimization allows us to conduct reasonably robust quantum simulations on current quantum hardware.}

	%========================================================FIG 7===================================================================

%==================================================================================================================================

Furthermore, when we implement our VQA circuits on the noisy IBM Q processor, we select a qubit layout on the physical device with the lowest gate errors possible, based on the latest calibration data and optimal compilation of the circuit. Here, our trained one-dimensional ansatz circuit only consists of CX gates and $U_{3}$ gates, which can be directly transpiled to the physical ``ibm$\_$toronto" device which uses the same set of basis gates.

\subsection{Readout error mitigation} A readout error occurs when the actual qubit to be read is spin-down (spin-up) but is read as spin-up (spin-down) instead. Since post-selection on the ancilla qubits is required to implement our non-unitary evolution, readout errors can lead to poor measurement results. Therefore, mitigating measurement errors~\cite{mthree} is an essential part of quantum simulations in the NISQ era. To achieve this, we perform readout error mitigation for all our circuit runs using a scalable approach known as \texttt{mthree} (M3)~\cite{mthree}.
Compared to the default methods provided by Qiskit~\cite{Qiskit}, which require $2^{L}$  circuits where $L$ is the number of qubits, \texttt{mthree} works in a subspace defined by the noisy bitstrings and only requires $2 \times L$ circuits for a single submission. This is integrated into our workflow of job submissions on the IBM Q devices (see Supplementary Information for job submission details). It is also capable of handling both uncorrelated and correlated readout errors on physical devices. 

\rz{In our work, the hardware simulations of FIG.~\ref{fig:nhsshmain} are conducted on the ``ibmq\_toronto" IBM quantum processor, and the hardware simulations of FIG.~\ref{fig:111000} and FIG.~\ref{fig:dynamics3} are performed on the ``ibmq\_brisbane" IBM quantum processor with both improved noise conditions and lower gate error}.
\subsection{Fermi skin profile from many-fermion NHSE}

\subsubsection{General formalism}
Here we show how a Fermi skin profile in-principle arises when the non-Hermitian skin effect (NHSE) and many-body fermionic exclusion are simultaneously present. Consider a complete orthonormal set of single-body right eigenstates $\{\ket{\phi_n^R}\}$, $n=1,...,L$ of an initially Hermitian Hamiltonian on $L$ sites, arranged in order of increasing eigenenergy. The simplest way to introduce NHSE is by means of a complex flux $\kappa$, which corresponds to the rescaling of hopping terms $t_{ij}c^\dagger_j c_i\rightarrow t_{ij}e^{-\kappa (x_i-x_j)}c^\dagger_j c_i$ between sites $i,j$ at positions $x_i,x_j$. This makes the hopping amplitude between sites $i$ and $j$ a factor of $e^{2\kappa(x_j-x_i)}$ times larger that of the reversed hopping from $j$ to $i$, and deforms the eigenstates according to $\phi_n^R(x)\rightarrow \bar\phi_n^R(x)=e^{-\kappa x}\phi_n^R(x)$. From now on, barred states will refer to NHSE-pumped states, with the superscript ``R'' emphasizing that they are right eigenstates.

Under this NHSE pumping, an $N$-fermion state (with $N\leq L$) can be constructed as the Slater determinant
\begin{eqnarray}
\bar \Psi_N^R(x_1,...,x_N) &=& \sum_{\sigma}(-1)^\sigma \prod^N_{n=1}e^{-\kappa x_n}\phi_{\sigma(n)}^R(x_n)\notag\\
 &=& e^{-\kappa \sum\limits_{j=1}^N x_j}\sum_{\sigma}(-1)^\sigma \prod^N_{n=1}\phi_{\sigma(n)}^R(x_n),\notag\\
\end{eqnarray}
where $\sigma$ is a permutation of indices $1,2,...,N$. As such, $\bar \Psi_N^R(x_1,...,x_N)$ is an antisymmetric linear combination of products of the $N\leq L$ occupied single-body eigenstates with position indices taken over all $N!$ possible permutations. While the $e^{-\kappa \sum_{j=1}^N x_j}$ term may seem to suggest that the $N$-fermion state still exhibits exponential spatial localization, this cannot be allowed by Pauli exclusion, as enforced by the antisymmetry of the determinant. To see this explicitly, we compute the spatial density (not biorthogonal density)
\begin{widetext}
\begin{eqnarray}
n_x&=&\langle\bar\Psi_N^R|c^\dagger_xc_x|\bar\Psi_N^R\rangle\notag\\
&=& \prod_{j=1}^{N-1}\sum_{x_j=1}^L|\bar\Psi_N^R(x_1,...,x_{N-1},x)|^2\notag\\
&=& e^{-2\kappa x}e^{-2\kappa \sum_{j=1}^{N-1} x_j}\prod_{j=1}^{N-1}\sum_{x_j=1}^L\left|\sum_{\sigma}(-1)^\sigma \phi_{\sigma(N)}^R(x)\prod^{N-1}_{n=1}\phi_{\sigma(n)}^R(x_n)\right|^2\notag\\
&=& e^{-2\kappa x}\sum_{\sigma,\sigma'}(-1)^{\sigma+\sigma'}\phi_{\sigma(N)}^{R*}(x)\phi_{\sigma'(N)}^R(x)\prod_{j=1}^{N-1}\left(\sum_{x_j=1}^Le^{-2\kappa x_j}\phi_{\sigma(j)}^{R*}(x_j)\phi_{\sigma'(j)}^{R}(x_{j})\right) \notag\\
&=& \sum_{m,n=1}^N(-1)^{m+n}\bar\phi_{m}^{R*}(x)\bar\phi_{n}^R(x)\sum_{\sigma_m,\sigma'_n}(-1)^{\sigma_m+\sigma'_n}\prod_{j=1}^{N-1}\langle\bar \phi^R_{\sigma_m(j)}|\bar \phi^R_{\sigma_n'(j)}\rangle \notag\\
&=& \sum_{m,n=1}^N\bar\phi_{m}^{R*}(x)\left[B^{-1}\right]_{mn}\bar\phi_{n}^R(x)
\label{nxB}
\end{eqnarray}
\end{widetext}
where $B_{mn}=\langle\bar \phi^R_{m}|\bar \phi^R_{n}\rangle=\sum\limits_{x=1}^Le^{-2\kappa x}\phi^{R*}_m(x)\phi^R_n(x)$ is the $N\times N$ overlap matrix between the occupied NHSE eigenstates. The penultimate line was obtained by labeling $m=\sigma(N)$ and $n=\sigma'(N)$, and defining permutations $\sigma_m,\sigma'_n$ as the permutations of $1,...,N$ that exclude elements $m$ and $n$ respectively. To obtain the last line, we note that the penultimate line is just the contraction of the basis NHSE eigenstates with the adjugate matrix~\cite{rubiano2023higher} of the overlap matrix $B$, whose elements are just the determinants of the overlap matrix with the $m$-th row and $n$-th column  deleted. Note that $n_x$ thus defined is invariant under constant rescalings of any constituent single-body eigenstate, and that its overall normalization is determined by the requirement that $n_x\rightarrow \sum_m |\phi_m^R(x)|^2$ in the Hermitian limit.

\begin{figure*}	
	\centering
	\includegraphics[width=.83\linewidth]{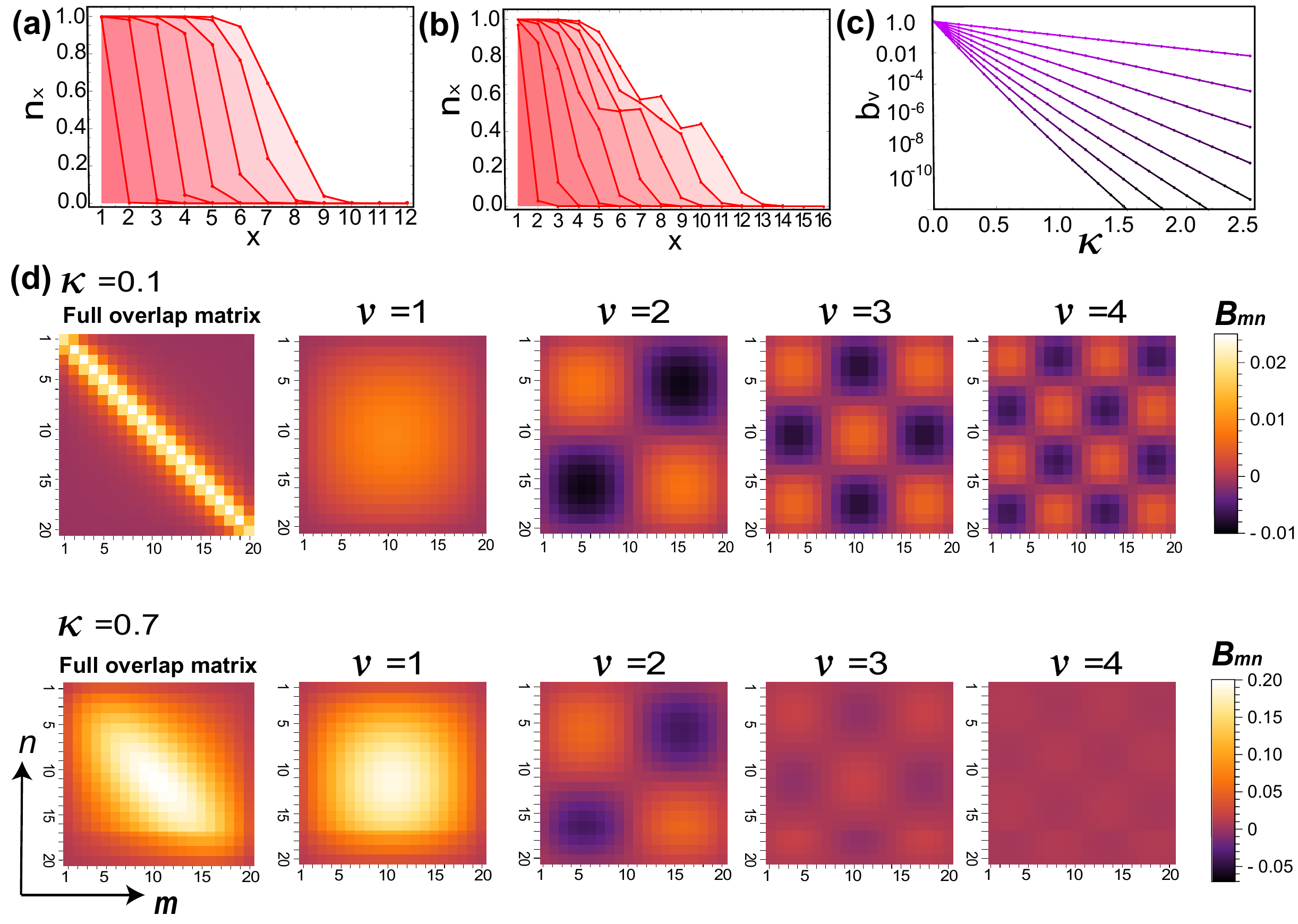}
		 %\includegraphics[width=.11\linewidth]{ortho_L20_oc20.pdf}
	 %\includegraphics[width=.11\linewidth]{ortho_L21_oc20.pdf}
		%\includegraphics[width=.11\linewidth]{ortho_L22_oc20.pdf}
		%\includegraphics[width=.11\linewidth]{ortho_L25_oc20.pdf}
		%\includegraphics[width=.11\linewidth]{ortho_L30_oc20.pdf}
		%\includegraphics[width=.11\linewidth]{ortho_L50_oc20.pdf}
		%\includegraphics[width=.11\linewidth]{ortho_L75_oc20.pdf}
		%\includegraphics[width=.11\linewidth]{ortho_L100_oc20.pdf}
		%\includegraphics[width=.05\linewidth]{ortho_legend.pdf}
	\begin{comment}
		\includegraphics[width=.32\linewidth]{nx_a10L12oc7.pdf}
	  \includegraphics[width=.32\linewidth]{nx_a5L16oc8.pdf}
	  \includegraphics[width=.33\linewidth]{bnu_L10oc8.pdf}\\
		\includegraphics[width=.21\linewidth]{overlap_L20_oc20_a11all.pdf}
		\includegraphics[width=.17\linewidth]{overlap_L20_oc20_a11_nu1.pdf}
		\includegraphics[width=.17\linewidth]{overlap_L20_oc20_a11_nu2.pdf}
		\includegraphics[width=.17\linewidth]{overlap_L20_oc20_a11_nu3.pdf}
		\includegraphics[width=.17\linewidth]{overlap_L20_oc20_a11_nu4.pdf}
		\includegraphics[width=.07\linewidth]{overlap_L20_oc20_a11legend.pdf}\\
		\includegraphics[width=.21\linewidth]{overlap_L20_oc20_a2all.pdf}
		\includegraphics[width=.17\linewidth]{overlap_L20_oc20_a2_nu1.pdf}
		\includegraphics[width=.17\linewidth]{overlap_L20_oc20_a2_nu2.pdf}
		\includegraphics[width=.17\linewidth]{overlap_L20_oc20_a2_nu3.pdf}
		\includegraphics[width=.17\linewidth]{overlap_L20_oc20_a2_nu4.pdf}
		\includegraphics[width=.07\linewidth]{overlap_L20_oc20_a2legend.pdf}
	\end{comment}
	\caption{\textbf{Orthogonalization of the $n_x$ profile into non-overlapping effective particle contributions and the eigen-structure of the overlap matrix $B_{mn}$.} (a,b) Plots of the many-body density profile $n_x=\sum_{\nu=1}^N  n_\nu(x)$ for (a) $N=7$, $L=8$, $\kappa = \log 10$ and (b) $N=8$, $L=16$, $\kappa = \log 5$ for the many-fermion Hatano-Nelson model with single-body eigenstates given in Eq.~\ref{phiHN}. The different colors show the accumulated partial contributions $\sum_{\nu=1}^{N_\text{partial}} n_\nu(x)$, where $N_\text{partial}=1,2,...,N$ in decreasing color intensity. (a) When the NHSE asymmetry $\kappa = \log 10$ is strong, $n_\nu(x)\approx \delta_{\nu x}$ for the first few $\nu$, leading to an almost flat $n_x$ profile at small $x$, and a Fermi-Dirac-like overall profile for general $x$. (b) For lower $\kappa$, this ``Fermi skin'' profile can become slightly non-monotonic.
(c) The eigenvalues $b_\nu$ of the overlap matrix $B$ are exponentially separated with intervals proportional to the NHSE asymmetry $\kappa$. For $\kappa>1$, the eigenvalues exhibit a clear hierarchy of scales. 
(d) The decomposition of the overlap matrix $B=\sum_\nu b_\nu \varphi_\nu\varphi_\nu^T$ of the right skin eigenstates in terms of the contributions from the $\nu$-th eigensector, for small hopping asymmetry $\kappa=0.1$ (Top) and larger asymmetry $\kappa=0.7$ (Bottom), both at $N=L=20$. $B$ is manifestly well-approximated by just its $\nu=1$ eigensector for large $\kappa$, consistent with the hierarchy of scales in the eigenspectrum of $B$.	
}
	\label{fig:nx}
\end{figure*}  

Eq.~\ref{nxB} has effectively expressed the many-body density $n_x$ entirely in terms of the single-body eigenstates $\bar \phi^R_n$ and their overlaps. In the Hermitian case without the NHSE, these eigenstates are already orthonormal and the overlap matrix is trivially the identity. The non-trivial profile of the many-body density is thus a direct consequence of the non-orthonormality of the NHSE-pumped (right) eigenstates. Further insight can be gleaned from expanding the overlap matrix $B$ in its eigenbasis: $B=\sum_\nu b_\nu \varphi_\nu\varphi_\nu^T$, whose eigenstates $\varphi_\nu$ are orthonormal and can be obtained from the $\phi^R_m$s by Gram-Schmidt orthogonalization. Then Eq.~\ref{nxB} can be expressed as
\begin{eqnarray}
n_x &=& \sum_{m,n=1}^N\bar\phi_{m}^{R*}(x)\left[\sum_\nu b_\nu^{-1} \varphi_\nu\varphi_\nu^T\right]_{mn}\bar\phi_{n}^R(x)\notag\\
&=&\sum_\nu b_\nu^{-1}|\varphi_\nu^T\cdot \bar \phi^R(x)|^2=\sum_\nu n_\nu(x)
\label{nxB2}
\end{eqnarray}
where $\bar\phi^R(x)=(\bar \phi^R_1(x),\bar \phi^R_2(x),...,\bar \phi^R_N(x))^T$, and $n_\nu(x)$ is the contribution to the many-body density $n_x$ from the $\nu$-th eigenstate $\varphi_\nu$ of the overlap matrix. Notably, $n_\nu(x)$ has the physical interpretation of being the density of the $\nu$-th effective NHSE-squeezed ``particle'' (see FIG.~\ref{fig:nx}a,b), since $\sum\limits_{x_j=1}^L n_\nu(x)=b_\nu^{-1}\varphi_\nu^T\cdot\left[\sum\limits_{x=1}^L\bar\phi^{R*}(x)\bar \phi^R(x) \right]\varphi_\nu=b_\nu^{-1}\varphi_\nu^T \cdot B\varphi_\nu=b_\nu^{-1}b_\nu \varphi_\nu^T\cdot \varphi_\nu=1$ is always normalized to unity, as expected of a bona-fide particle.

The spectral decomposition of the overlap matrix in Eq.~\ref{nxB2} allows one to mathematically deduce the mathematical origin of the Fermi skin profile, applicable for generic multi-fermion systems with sufficiently large NHSE. From physical grounds, we already expect the fermions to accumulate against the left boundary at $x=0$, even though $n_x\leq 1$ due to Pauli exclusion. Below we explain how, from the non-orthonormality and incompleteness of the set of $\ket{\bar\phi^R_n}$, $n=1,...,N$ eigenstates, we can deduce that as we move away from the left boundary, the $n_x$ profile evolves from a near-uniform value of $n_x\approx 1$ for $x\ll N$, and gradually decays to $n_x\approx 0$ for $x\gg N$ when $N<L$ (FIG.~\ref{fig:nx}a,b).

We first express the overlap matrix as
\begin{equation}
B= \sum\limits_{\nu=1}^N b_\nu \varphi_\nu\varphi_\nu^T=\sum\limits_{x=1}^Le^{-2\kappa x}\phi^R(x)[\phi^R(x)]^T.
\label{nxB3}
\end{equation}
In the fully occupied $N=L$ case, we could instantly identify $b_\nu = e^{-2\kappa\nu}$ and $\varphi_\nu=\phi^R(\nu)$, such that $n_\nu(x)= e^{2\kappa\nu}|[\bar\phi^{R*}(\nu)]^T\cdot\bar\phi^R(x)|^2=|[\phi^{R*}(\nu)]^T\cdot\phi^R(x)|^2=\delta_{\nu x}$ due to the completeness of the (Hermitian) $\phi^R_n$ eigenstates.

When the fermionic occupancy $N<L$ is not full, $n_\nu(x)\neq \delta_{\nu x}$ because the $N$ occupied $\phi^R_n$ are no longer complete. However, in the strong NHSE limit where $e^{-\kappa}\ll 1$, the $e^{-2\kappa x}$ factor in Eq.~\ref{nxB3} still forms a well-defined hierarchy of scales with exponential spacing (FIG.~\ref{fig:nx}c) that prevent easy mixing of the states $\phi^R(x)$~\footnote{In particular, the overlap matrix $B_{mn}$ is dominated by a rank-1 subspace of largest eigenvalues because $B_{mn}=\sum\limits_{x=1}^Le^{-2\kappa x}\phi^{R*}_m(x)\phi^R_n(x)\approx  \phi^{R*}_m(1)\phi^R_n(1) $.}. Hence we still have $n_\nu(x)\approx \delta_{\nu x}$, at least for the first few $\nu$, which results in an almost uniform $n_x\approx \sum\limits_{\text{small }\nu}\delta_{\nu x}$ profile at small $x$ (first few ``slices'' in FIG.~\ref{fig:nx}a,b). However, as $\nu$ increases, obtaining the orthonormal eigenstate $\varphi_\nu$ requires more and more mixing from the $\phi^R$ at various sites $x$. As shown in FIG.~\ref{fig:nx}a,b, when $x$ and $\nu\approx N$, $n_\nu(x)$ has spread out significantly (light-colored slices), and gives way to the rapid drop of $n_x$ in the ``Fermi skin'' profile. 

Interestingly, even though the overlap matrix $B$ is dominated by the leading $\nu=1$ eigenvalue contribution $b_1\varphi_1\varphi_1^T$ at large $\kappa$ (FIG.~\ref{fig:nx}d), all eigensectors $\nu$ in fact contribute {\it equally} to density profile $n_x=\sum_\nu n_\nu(x)$. This is because each constituent $n_\nu(x)$ is in fact proportional to $b_\nu^{-1}$ (Eq.~\ref{nxB2}), such that they obey the same normalization $\sum_{x=1}^L n_\nu(x)=1$.

\begin{comment}
The above formalism can be directly generalized to obtain many-body $\mathcal{N}$-point functions $\langle\bar\Psi_N^R|\Pi\limits_{j=1}^\mathcal{N}c^\dagger_{x_j}c_{x_j}|\bar\Psi_N^R\rangle$. We replace the overlap matrix $B_{mn}$ in Eq.~\ref{nxB} by an $^NC_\mathcal{N}\times ^NC_\mathcal{N}$ matrix whose elements are the generalized overlaps $\Pi\limits_{j=1}^n \bar\phi^R_n$ between all $^NC_\mathcal{N}$ unordered products of $\mathcal{N}$ different single-particle eigenfunctions.
.........
$\bar \phi_m^R,\bar \phi_n^R$ appearing in Eq.~\ref{nxB}, we 
\end{comment}

\subsubsection{Fermi skin for the Hatano-Nelson model}
Below, we present details of the Fermi skin profile for the analytically tractable Hatano-Nelson (HN) eigenstates, even though our formalism is completely applicable for generic single-body eigenstates, such as those of the non-Hermitian SSH model appearing in the text.

In an OBC HN chain with $L$ sites, a complete orthonormal basis (in the Hermitian limit) is given by
\begin{equation}
\phi^\text{HN}_n(x) = \sqrt{\frac{2}{L+1}}\sin\frac{\pi nx}{L+1}
\label{phiHN}
\end{equation}
where we have dropped the ``R'' superscript. Introducing the NHSE by rescaling the left/right hoppings by a factor of $e^{\pm \kappa}$, these eigenstates are deformed into $\bar\phi^\text{HN}_n(x)=e^{-\kappa x}\phi^\text{HN}_n(x)$ right NHSE eigenstates.

For this simplest model, an analytic expression also exists for the overlap matrix elements  $B^\text{HN}_{mn}=\langle\bar \phi^\text{HN}_{m}|\bar \phi^\text{HN}_{n}\rangle=$ 
\begin{equation}
\frac{\left(1-(-1)^{m+n}e^{-2\kappa(L+1)}\right)\sinh 2\kappa \sin \frac{m\pi}{L+1}\sin\frac{n\pi}{L+1}}{(L+1)\left(\cosh 2\kappa-\cos\frac{\pi(m+n)}{L+1}\right)\left(\cosh 2\kappa-\cos\frac{\pi(m-n)}{L+1}\right)},
\end{equation}
as plotted in FIG.~\ref{fig:nx}d for small $\kappa=0.1$ (Top) and moderate $\kappa=0.7$ (Bottom). In the $e^\kappa\gg 1 $ limit of strong NHSE, we are indeed left with the dominant rank-one expression $B^\text{HN}_{mn}\rightarrow \frac{2e^{-2\kappa}}{L+1} \sin \frac{m\pi}{L+1}\sin\frac{n\pi}{L+1}$. By explicit fitting of the resultant $n_x$ with the Fermi skin ansatz $n_\text{FS}(x)=\left(1+e^{\beta_\text{eff}(x-N)}\right)^{-1}$ (Eq.~\ref{nFs}), which holds accurately in the large $e^\kappa$ limit, we obtain the approximate linear relationship $\beta_\text{eff}\approx 4\kappa$ for this HN model~\cite{mu2020emergent}. This is expected to hold for generic models too, as long as $\kappa$ is sufficiently large such that $n_x$ is monotonic.

\subsubsection{Observing the Fermi skin from short-time evolution}

Technically, the many-body density should only accurately assume the Fermi skin profile $n_\text{Fs}(i)$ (Eq.~\ref{nFs}) in the asymptotic limit of long times and large system sizes. Yet, we are able to observe an approximate Fermi skin-like density profile (Fig.~\ref{fig:111000}) in only 5 to 10 Trotter steps. 
%That so few Trotter steps can already allow for the observation of an approximately Fermi skin density profile is possible 
Below, we provide clues to why this is possible by considering the following bespoke measure of fermion density, which is weighted by the overlap of the initial state $\ket{\psi(0)}$ with the many-body eigenstates $\ket{\psi_{j}}$:  
\begin{equation}\label{statistics}
	n^{\prime}(i)=C\sum_{j}|\bra{\psi(0)}\ket{\psi_{j}}|(\bra{\psi_{j}}\hat{Z}_{i}\ket{\psi_{j}}+1)/2,
\end{equation}
where $C$ is a normalization factor that ensures the half-filled condition $\sum_{i=0}^{L-1} n'(i)=L/2$.  We here effectively extract the contribution from high-overlap states by adopting $|\bra{\psi(0)}\ket{\psi_{j}}|$ as weights. Thus, $n^{\prime}(i)$ serves as an estimate for the averaged spatial density profile at short times since short-time dynamics are governed by eigenstates with high initial overlaps.

\begin{figure}
	\centering
	\includegraphics[width=0.99\linewidth]{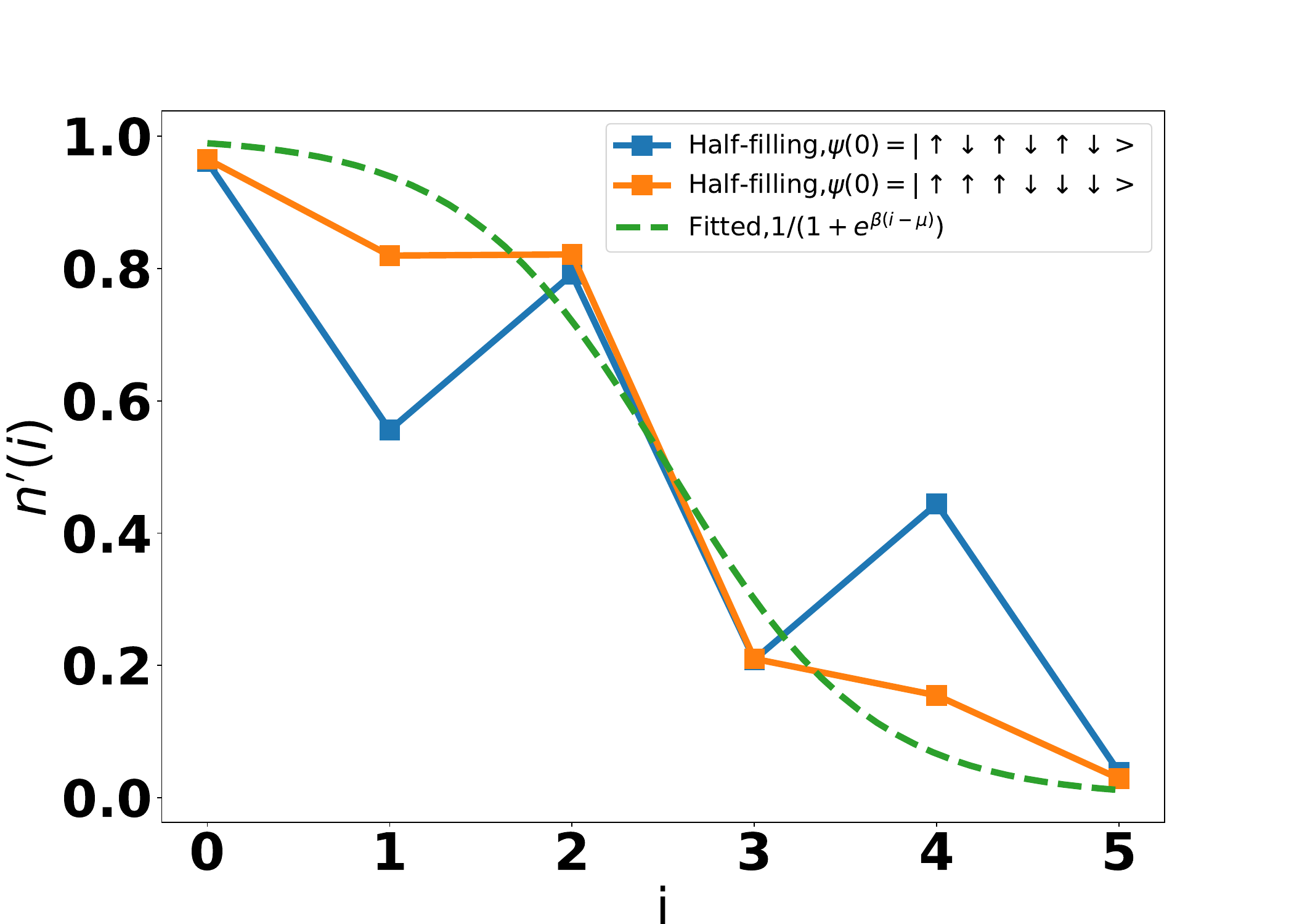}
	\caption{\textbf{Illustration of the $n^{\prime}(i)$ for chosen initial states.} The numerically evaluated $n^{\prime}(i)$ from Eq.~\ref{statistics} for our model Eq.~\ref{spinssh_main} ($J=2,\gamma=1.5$), with initial states $\ket{\uparrow\downarrow\uparrow\downarrow\uparrow\downarrow}$ and $\ket{\uparrow\uparrow\uparrow\downarrow\downarrow\downarrow}$ represented in blue and yellow respectively. The yellow profile exhibits a better fit with the Fermi-Dirac profile $1/(1+e^{\beta(i-\mu)})$ (green), where $\mu\approx(L-1)/2=2.5$ and $\beta=\ln(\frac{J+\gamma}{J-\gamma})\approx1.95$ is empirically optimized.}
	\label{fig:fig6}
\end{figure}

As shown in FIG.~\ref{fig:fig6}, we compare the $n^{\prime}(i)$  of two different initial states: the density-wave (antiferromagnetic) state ($\psi(0)=\ket{\uparrow\downarrow\uparrow\downarrow\uparrow\downarrow}$, blue) and the half-localized state ($\psi(0)=\ket{\uparrow\uparrow\uparrow\downarrow\downarrow\downarrow}$, yellow) used in our hardware demonstration of the Fermi skin. While the half-localized initial state (yellow curve) approximately exhibits a Fermi-Dirac real-space profile, as seen from its fit with the green dashed curve, the antiferromagnetic initial state does not.  Hence had we used the latter as the initial state for demonstrating multi-fermion NHSE, we would have had to execute more Trotter steps before it converges approximately to the Fermi skin profile exhibiting its characteristic Fermi-Dirac shape.

\newpage
%%%%%%%%%%%%%%%%%%%%%%  References %%%%%%%%%%%%%%%%%%%%%%
%\bibliographystyle{apsrev4-1}
%\bibliographystyle{plain}
%\bibliography{references}
\bibliography{ref}
\setcounter{equation}{0}
\setcounter{figure}{0}
\setcounter{table}{0}
\setcounter{section}{0}
\renewcommand{\theequation}{S\arabic{equation}}
\renewcommand{\thefigure}{S\arabic{figure}}
\renewcommand{\thesection}{S\arabic{section}}
\renewcommand{\thepage}{S\arabic{page}}
\onecolumngrid
\flushbottom
\newpage
\appendix

\onecolumngrid

\setcounter{equation}{0}
\setcounter{figure}{0}
\setcounter{table}{0}
\setcounter{section}{0}
\renewcommand{\theequation}{S\arabic{equation}}
\renewcommand{\thefigure}{S\arabic{figure}}
\renewcommand{\thesection}{S\arabic{section}}
\renewcommand{\thepage}{S\arabic{page}}

\appendix
\subsection*{\normalsize Supplementary Information for ``Observation of the non-Hermitian skin effect on a digital quantum computer''}

\maketitle
\section{S1. Details of implementing non-unitary evolution via post-selection}

Below, we provided a detailed description of the ancilla-based quantum circuit that implements the non-unitary dynamics for our two models in the text. Their accuracy is benchmarked against exact diagonalization results in the absence of device noise. Some formalisms have been repeated from the text in order to make this section self-contained.

\subsection{Non-Hermitian Su-Schrieffer–Heeger model}

We consider a non-Hermitian Su-Schrieffer–Heeger (nH-SSH) fermionic chain of length $L$ under open boundary conditions (OBCs) 
\begin{equation}\label{ssh}
	\hat{H}^{\rm Fermion}_{\rm nH-SSH}=-\sum_{j=0}^{L/2}\left((J+\gamma)\hat{f}_{2j}^{\dagger} \hat{f}_{2j+1}+(J-\gamma)\hat{f}_{2j+1}^{\dagger} \hat{f}_{2j}\right)-\sum_{j=0}^{L/2-1}\left(J\hat{f}_{2j+2}^{\dagger} \hat{f}_{2j+1}+J\hat{f}_{2j+1}^{\dagger} \hat{f}_{2j+2}\right).
\end{equation}
Through the Jordan-Wigner transformation~\cite{coleman2015introduction} which maps between fermions and spins, we obtain an equivalent non-Hermitian spin chain from Eq.~\ref{ssh} as 
\begin{equation}\label{spinssh}
	\begin{aligned}
%		\hat{H}^{\rm Spin}_{\rm nH-SSH}=-\frac{J}{2}\sum_{j=0}^{L-2}(\hat{X}_{j}\hat{X}_{j+1}+\hat{Y}_{j}\hat{Y}_{j+1})-\frac{i\gamma}{2}\sum_{j=0}^{L/2}(\hat{Y}_{2j}\hat{X}_{2j+1}-\hat{X}_{2j}\hat{Y}_{2j+1}),\\
		\hat{H}^{\rm Spin}_{\rm nH-SSH}=-\sum_{j=0}^{L/2}\left((J+\gamma)\hat{X}^{+}_{2j}\hat{X}^{-}_{2j+1}+(J-\gamma)\hat{X}^{-}_{2j}\!\hat{X}^{+}_{2j+1}\right)-\frac{J}{2}\sum_{j=0}^{L/2-1}\left(\hat{X}_{2j+1}\hat{X}_{2j+2}+\hat{Y}_{2j+1}\hat{Y}_{2j+2}\right).
	\end{aligned}
\end{equation}
with $\hat{X}^{\pm}=\frac{\hat{X}\pm i\hat{Y}}{2}$, $\hat X, \hat Y$ Pauli matrix operators. Fortuitously, the nearest-neighbor fermionic couplings also map to nearest-neighbor spin interactions, despite the non-locality of the Jordan-Wigner transformation in general. \rz{By virtue of the Jordan-Wigner transformation, which preserves fermionic operator anticommutativity, the spins in this spin chain now describe effective fermions that obey the Pauli exclusion principle.}

To simulate the dynamics for Eq.~\ref{spinssh} on a quantum circuit, we employ our ``local'' approach by breaking up the time evolution into $T/\delta t$ Trotter steps, each of duration $\delta t$, such that each step is approximated by two separate exponentials:
\begin{equation}\label{suppdy}
	\begin{aligned}
		e^{-iT \hat{H}^{\rm Spin}_{\rm nH-SSH}}&=(e^{-i\delta t \hat{H}^{\rm Spin}_{\rm nH-SSH}})^{\frac{T}{\delta t}}\\
		&\approx ([\prod_{j}e^{i\delta t(+\frac{J}{2}(\hat{X}_{2j+1}\hat{X}_{2j+2}+\hat{Y}_{2j+1}\hat{Y}_{2j+2}))}] [\prod_{j}e^{+i\delta t((J+\gamma)\hat{X}^{+}_{2j}\hat{X}^{-}_{2j+1}+(J-\gamma)\hat{X}^{-}_{2j}\!\otimes\!\hat{X}^{+}_{2j+1})}])^{\frac{T}{\delta t}}.
	\end{aligned}
\end{equation}
Every Trotter step contains the two-qubit non-unitary operator
\begin{equation}\label{xy}
	R^{\prime\prime}=e^{+i\delta t( (J-\gamma)\hat{X}^{+}_{j+1}\!\!\hat{X}^{-}_{j}+(J+\gamma)\hat{X}^{-}_{j+1}\!\!\hat{X}^{+}_{j})},
\end{equation}
which can be implemented in our quantum circuit with the help of an additional ancilla qubit, as explained in the text.
This is done by embedding $R''$ into the larger unitary $U_{R^{\prime\prime}}$
\begin{equation}\label{rssh}
	U_{R^{\prime\prime}}=\left[\begin{array}{cc}
		uR^{\prime\prime} & B_{4\times4} \\                                                                   
		C_{4\times4} & D_{4\times4}
	\end{array}\right].
\end{equation}
where $u^{-2}$ is the maximum eigenvalue of $R^{\prime\prime\dagger}R^{\prime\prime}$. More details of solving $B_{4\times4}$, $C_{4\times4}$, and $D_{4\times4}$ are described in the Methods. To recover the action of the original non-unitary $R''$ in the quantum circuit requires the post-selection of $\ket{\uparrow}$ on every ancilla qubit. For realizing the dynamics of the nH-SSH model of length $L$, we require $L/2$ $U_{R^{\prime\prime}}$ operators in every Trotter step.

\begin{figure}[h]
	\centering
	\includegraphics[width=0.75\linewidth]{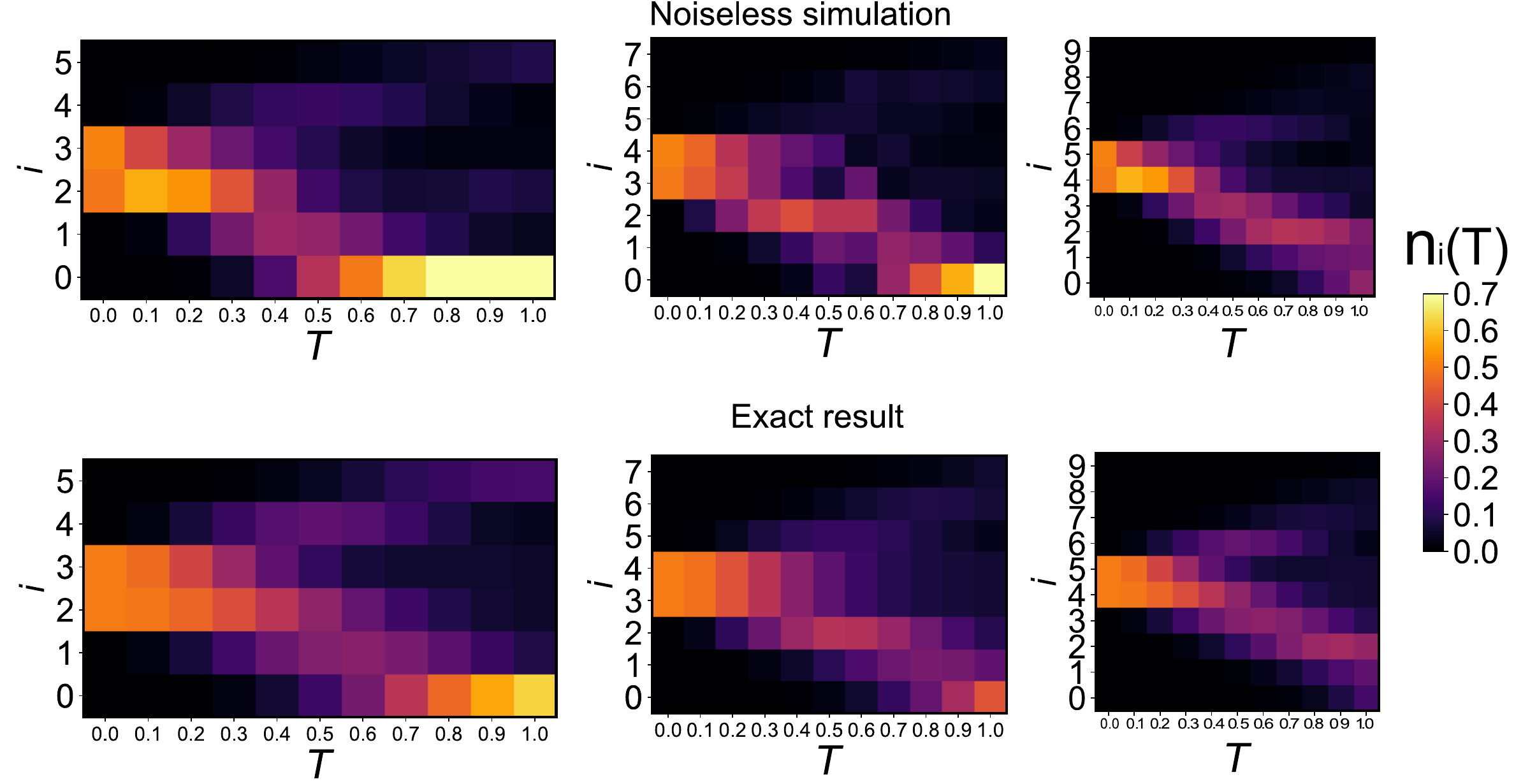}
	\caption{The good agreement between the exact diagonalization and noiseless circuit simulation results for the dynamical density evolution $n_{i}(T)=(\bra{\psi(T)} \hat{Z}_{i}\ket{\psi(T)}+1)/2$ for the non-Hermitian SSH model with a  single fermion, as governed by Eq.~\ref{spinssh}.  The single-fermion initial states are prepared as $(\ket{\downarrow\downarrow\downarrow\uparrow\downarrow\downarrow}+\ket{\downarrow\downarrow\uparrow\downarrow\downarrow\downarrow})/\sqrt{2}$, $(\ket{\downarrow\downarrow\downarrow\downarrow\uparrow\downarrow\downarrow\downarrow}+\ket{\downarrow\downarrow\downarrow\uparrow\downarrow\downarrow\downarrow\downarrow})/\sqrt{2}$, and $(\ket{\downarrow\downarrow\downarrow\downarrow\downarrow\uparrow\downarrow\downarrow\downarrow\downarrow}+\ket{\downarrow\downarrow\downarrow\downarrow\uparrow\downarrow\downarrow\downarrow\downarrow\downarrow})/\sqrt{2}$ for $L=6,8,10$. Clear signatures of asymmetric NHSE propagation are evident whenever the non-Hermiticity $\gamma=0.5$. The noiseless simulation is computed on the Qiskit Aer simulator. We set $J=2.0$ with $\delta t=0.1$, and 10 steps with 10000 shots for each execution.
	}
	\label{fig:nhssh}
\end{figure}

The accuracy of our Trotterization and post-selection approach is evident in Figs.~\ref{fig:nhssh}, which compares the time evolution under different system sizes from a noiseless quantum circuit simulator with that from exact diagonalization. Here, we demonstrate the effect of post-selection in noiseless simulations. As we increase the system size, there always remains a good agreement between exact results and noiseless simulations. This fact indicates that the discrepancy of our hardware-run results in the main text is due to device noise (which is partly mitigated by the VQA circuit optimization), rather than the intrinsic inaccuracy of our approach.

\newpage
\subsection{Hatano-Nelson model}
For the non-Hermitian Hatano-Nelson model
\begin{equation}\label{HNsupp}
	\begin{aligned}
		\hat{H}^{\rm Spin}_{\rm HN}=-\sum_{j=0}^{L-1}\left((J+\gamma)\hat{X}^{+}_{j}\hat{X}^{-}_{j+1}+(J-\gamma)\hat{X}^{-}_{j}\!\hat{X}^{+}_{j+1}\right),
	\end{aligned}
\end{equation}
the dynamics for its 6-qubit chain in our work can be approximated in a quantum circuit through the Trotterization
\begin{equation}\label{HNdysupp}
	\begin{aligned}
		e^{-i\delta t\hat{H}^{\rm Spin}_{\rm HN}}\approx [U^{\rm nonH}_{0}U^{\rm nonH}_{2}U^{\rm nonH}_{4}][U^{\rm nonH}_{1}U^{\rm nonH}_{3}],
	\end{aligned}
\end{equation}
with 
\begin{equation}
	U^{\rm nonH}_{j}=e^{+i\delta t( (J-\gamma)\hat{X}^{+}_{j+1}\!\otimes\!\hat{X}^{-}_{j}+(J+\gamma)\hat{X}^{-}_{j+1}\!\otimes\!\hat{X}^{+}_{j})}.
\end{equation}
To realize the above non-unitary dynamics in a quantum circuit, we employ the ``global'' approach by directly embedding the 6-qubit operator $R_{\rm HN}=[U^{\rm nonH}_{0}U^{\rm nonH}_{2}U^{\rm nonH}_{4}][U^{\rm nonH}_{1}U^{\rm nonH}_{3}]$ into the following unitary 7-qubit operator
\begin{equation}
	U_{\rm HN}=\left[\begin{array}{cc}
		uR_{\rm HN} & B_{64\times64} \\		C_{64\times64} & D_{64\times64}
	\end{array}\right],
\end{equation}
where $U_{\rm HN}$ can be solved by the same method used in dealing with Eq.~\ref{rssh}. Combining all the Trotter steps,  the dynamical evolution operator for the Hatano-Nelson model is given by
\begin{equation}\label{HNdysupp2}
	\begin{aligned}
		e^{-iT\hat{H}^{\rm Spin}_{\rm HN}}\approx [U_{\rm HN}]^{T/\delta t}.
	\end{aligned}
\end{equation}
Note that the circuit for Eq.~\ref{HNdysupp2} only requires a single ancilla qubit, which avoids significant statistical errors from repeated post-selection in the noisy simulation.

\begin{figure}
	\centering
	\includegraphics[width=0.7\linewidth]{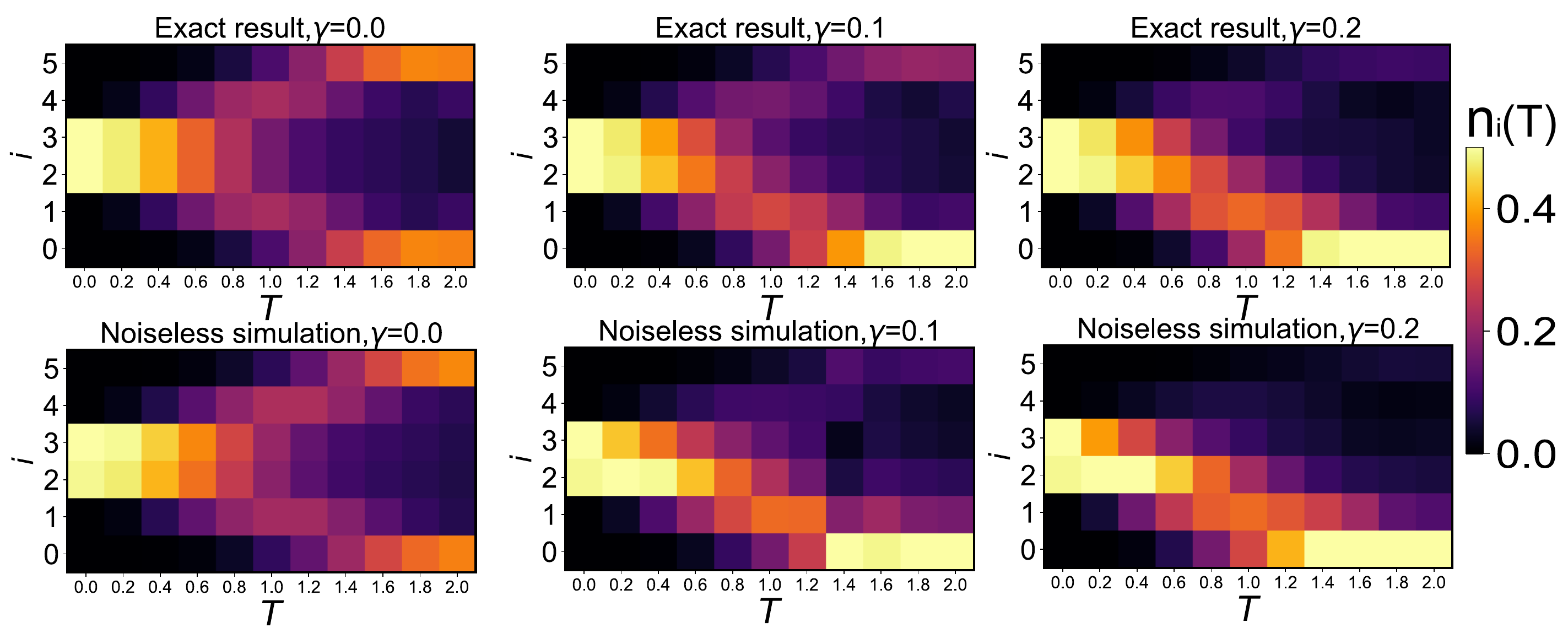}
	\caption{Almost perfect agreement between the exact and noiseless simulation results of the dynamical density $n_{i}(T)=(\bra{\psi(T)} \hat{Z}_{i}\ket{\psi(T)}+1)/2$ governed by our Hatano-Nelson Hamiltonian (Eq.~\ref{HNsupp}). With increasing non-Hermiticity $\gamma$, the NHSE evidently pumps the state towards qubit $i=0$. The noiseless simulations are implemented by Eq.~\ref{HNdysupp2} on the Qiskit Aer simulator, where the operator $[U_{\rm HN}]^{T/\delta t}$ is obtained by numerical decomposition. The initial state shown is the symmetrically centered $(\ket{\downarrow\downarrow\downarrow\uparrow\downarrow\downarrow}+\ket{\downarrow\downarrow\uparrow\downarrow\downarrow\downarrow})/\sqrt{2}$, and the initial ancilla qubit is $\ket{\uparrow}$. We set $J=2.0$ with $\delta t=0.1$, and 10 steps with 10000 shots for each execution.}
	\label{fig:nhhnsupp}
\end{figure}

\newpage
\section{S2. Details on the IBM Q device implementation}
\subsection{IBM Q devices}
Here, we present the details of the job submission and gate fidelities on IBM Q devices. Since the number of qubits for all the devices used on IBM Q is larger than the number of qubits ($\mathcal{L}=9$ is our entire system size including both physical and ancilla qubits) that we require, we have the liberty of first choosing good-quality qubits for our runs. Therefore, we first transpile the quantum circuits according to the geometry of each device, taking into account the single- and two-qubit gate errors with respect to the latest calibration data. As soon as the transpilation is completed, we record the indices of the qubits chosen from the device as a list of length $\mathcal{L}=9$ on the classical computer.

Then, we consider the readout error mitigation using the approach introduced in Ref.~\cite{mthree}. For each qubit, there are two outcomes $\uparrow$ and  $\downarrow$. Therefore, we need $2 \times\mathcal{L}=18$ calibration circuits. Following these, we execute both the transpiled circuits and calibration circuits on IBM Q quantum computers. This procedure is designed to maximize the readout error mitigation: the time interval between two consecutive executions is so short that the outcomes of the calibration circuits accurately reflect the noise of the transpiled circuits. Also, it suppresses stochastic noise as much as possible.

\begin{figure}
	\centering
	\includegraphics[width=0.8\linewidth]{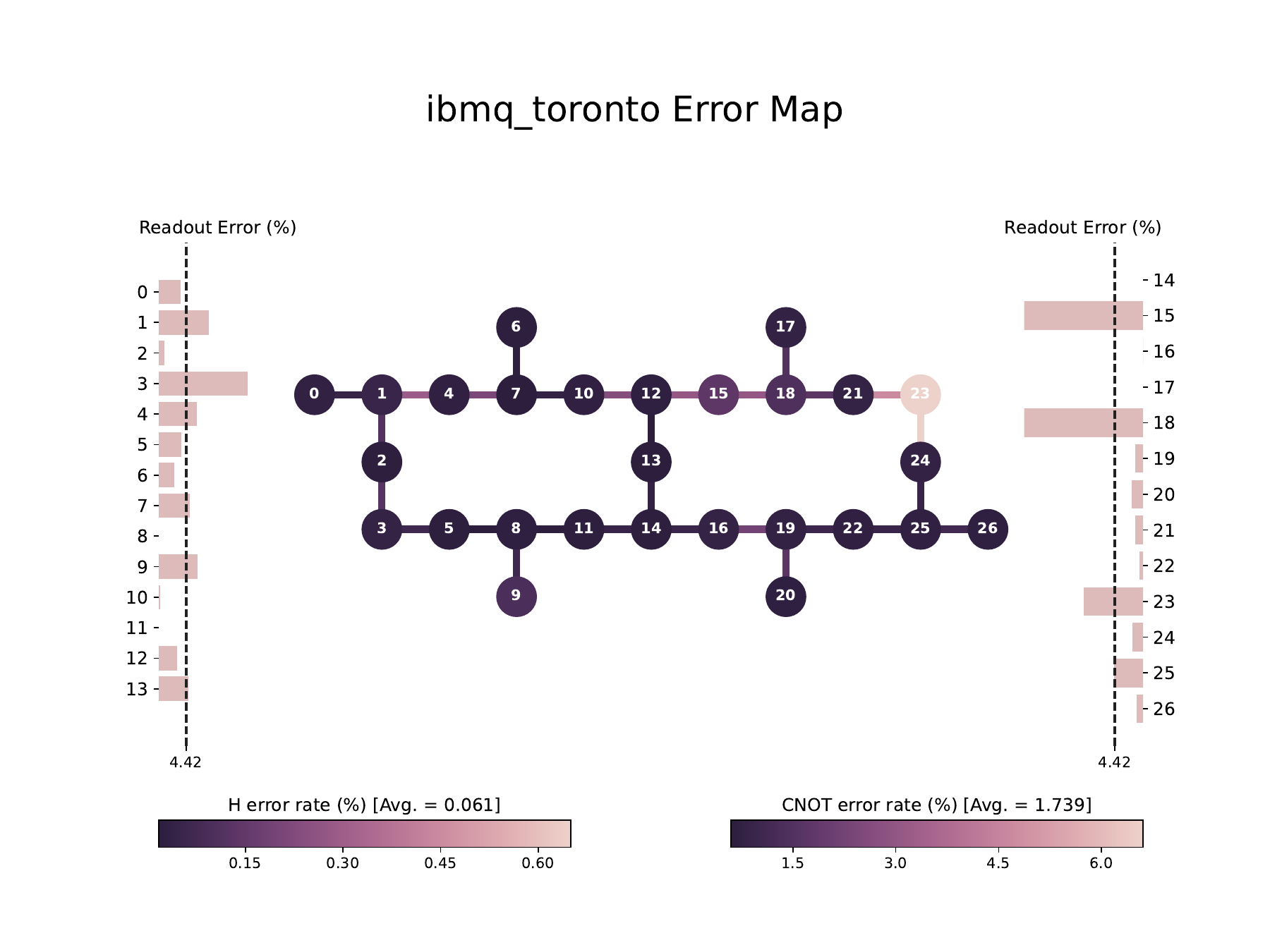}
	\caption{Details of hardware error specifications on the ``ibmq\_toronto" IBM quantum processor used for our hardware runs. The linear sequence of qubits $[1,2,3,5,8,11,14,13,12]$ is chosen for optimally avoiding the noisest gates and qubits with the highest readout errors, as explained in the calibration procedure above.}
	\label{fig:ibmq_device_supp}
\end{figure}

\begin{figure}
	\centering
	\includegraphics[width=1\linewidth]{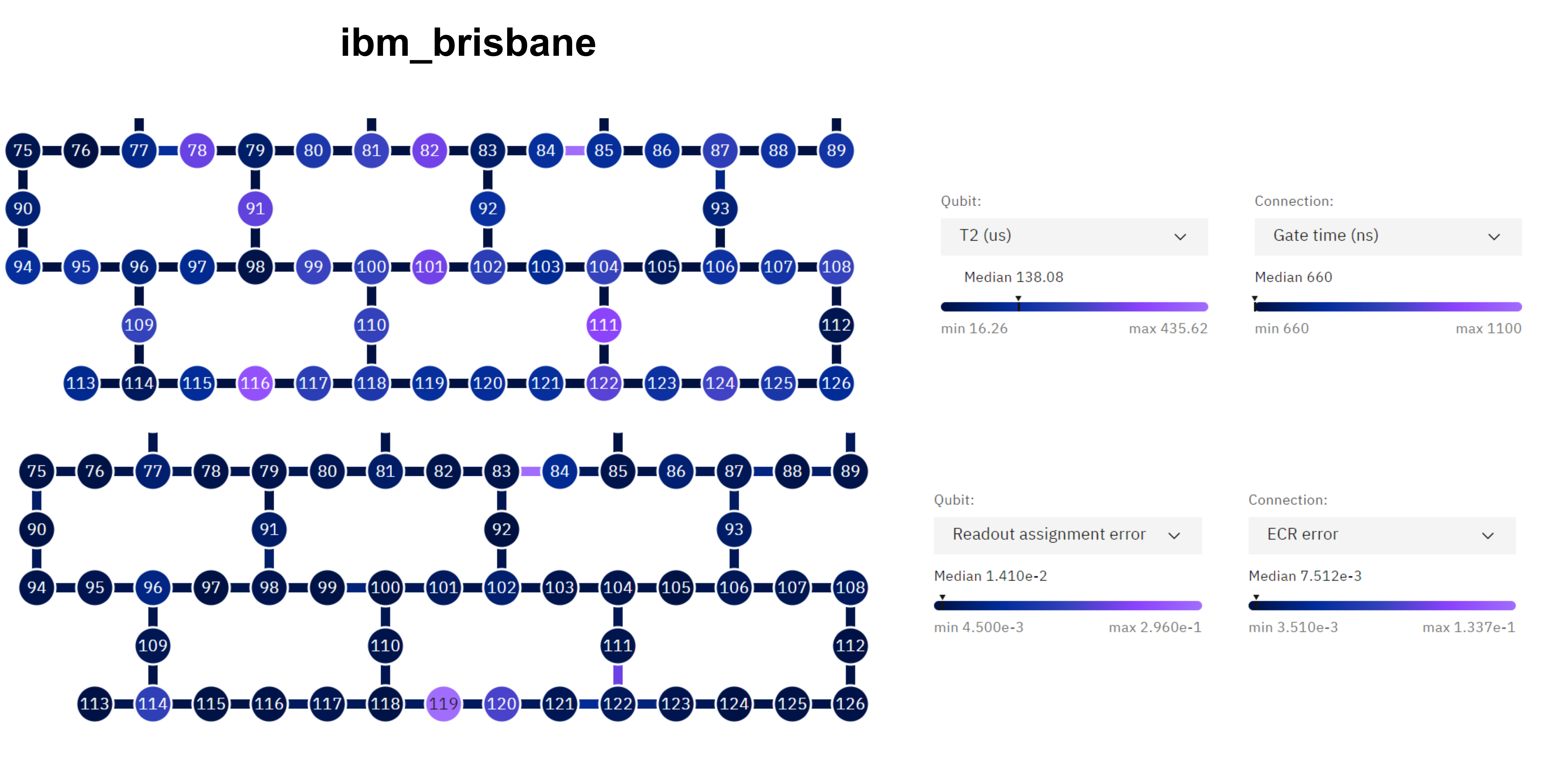}
	\caption{\rz{Details of hardware error specifications on the improved ``ibmq\_brisbane" IBM quantum processor. We provide information of $T_{2}$, gate time, readout errors, and Echoed Cross-Resonance (ECR) gate errors. The ECR gate is a maximally entangling gate and is equivalent to a CNOT gate up to single-qubit pre-rotations. Notably, this ECR gate demonstrates a lower error rate compared to the CNOT gate shown in FIG.~\ref{fig:ibmq_device_supp}. The simulation results shown in FIGs.~$4$, $5$ (main text), and FIG.~\ref{fig:dynamics} are executed on this hardware}. }
	\label{fig:ibm}
\end{figure}

\rz{In our work, we employ two different quantum devices: the first depicted in FIG.~\ref{fig:ibmq_device_supp} for  simulations in FIG.~$3$, and the second shown in FIG.~\ref{fig:ibm} for simulations in FIGs.~$4$,$5$. The advanced processor shown in FIG.~\ref{fig:ibm} offers higher-quality two-qubit gates. These enhancements lead to the improved results presented in  FIGs.~$4$,$5$.}

\newpage
\subsection{Variational quantum algorithms}
%\begin{figure}
%	\centering
%	\includegraphics[width=0.7\linewidth]{scaling}
%	\caption{}
%	\label{fig:scaling}
%\end{figure}

\begin{figure}
	\centering
	\includegraphics[width=0.8\linewidth]{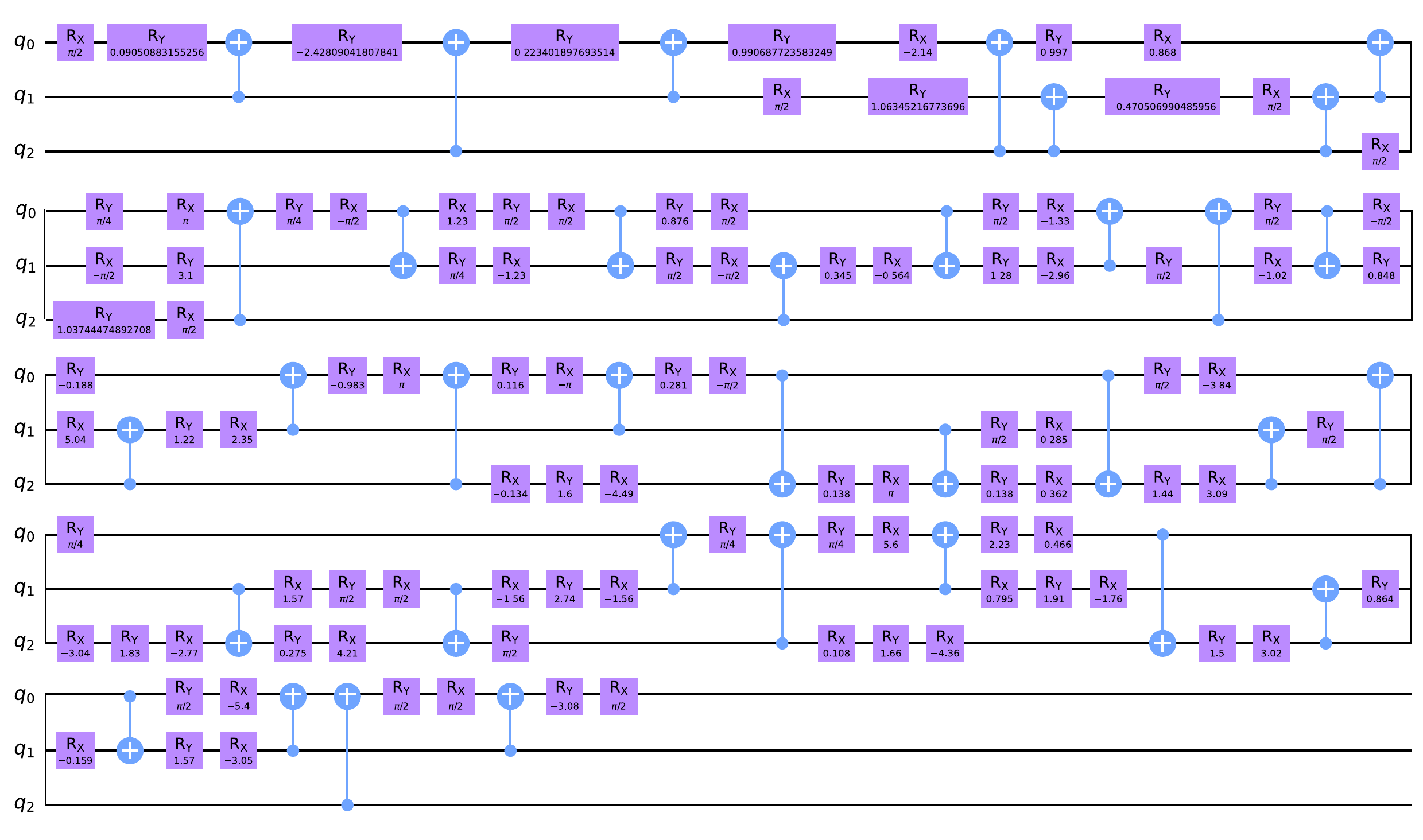}
	\caption{Details of the transpilation of the 3-qubit unitary Eq.~\ref{rssh} into basis gates, with $\delta t=0.1$, $J=2$, and $\gamma=0.15$. Here, the basis gates are $R_x$, $R_y$, and $CX$ gates. For each unit cell, $q_{1}$ and $q_{2}$ are the physical qubits, and $q_{0}$ is the ancilla qubit. Due to the considerable number of $CX$ gates, this decomposition is not ideal for simulations on actual noisy quantum devices.}
	\label{fig:transpileu}
\end{figure}

In FIG.~\ref{fig:ibmq_device_supp}, we show the details of gate error and the qubit indices for the IBM Q processor used. Both single-qubit Hadamard $H$ gates as well as the $CX$ gates are included. As shown, $CX$ gates constitute the most significant source of noise errors. Hence to limit hardware noise, our variational ansatz discussed in the text requires fewer CX gates than pure numerical circuit decomposition. An example of the numerical decomposition of the three-qubit operation in Eq.~\ref{xy} in terms of the basis gates is shown in FIG.~\ref{fig:transpileu}. 

%CH: Should probably not mention OBCs vs PBCs, because people would ask why we don't use the loops in the quantum processor
%Since the existing challenges with building circuits in ring layouts, i.e., coupling edges of a loop under periodic boundary conditions (PBCs) in current device topologies, our work focuses on the simulation of circuits under open boundary conditions  (OBCs)

\newpage
\subsection{Additional IBM Q simulations with different numbers of qubits}
\begin{figure}
	\centering
	\includegraphics[width=0.99\linewidth]{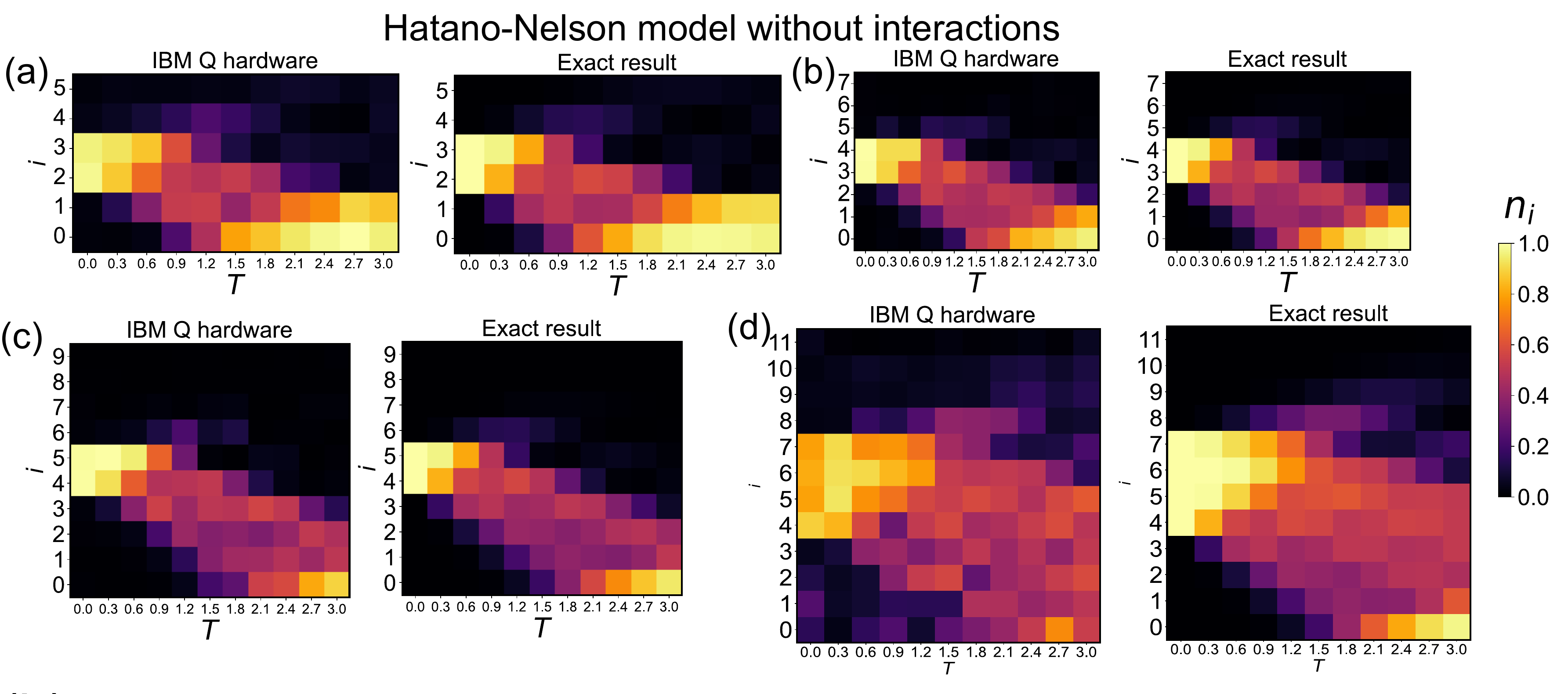}
	\caption{\rz{IBM Q hardware simulations of non-interacting Hatano-Nelson model (Eq.~\ref{HNsupp}) for different physical system sizes ($6$ to $12$ qubits) on  the improved ``ibmq\_brisbane" IBM quantum processor. For (a) to (d), initial states are: $\ket{\downarrow\downarrow\uparrow\uparrow\downarrow\downarrow}$, $\ket{\downarrow\downarrow\downarrow\uparrow\uparrow\downarrow\downarrow\downarrow}$, $\ket{\downarrow\downarrow\downarrow\downarrow\uparrow\uparrow\downarrow\downarrow\downarrow\downarrow}$, and $\ket{\downarrow\downarrow\downarrow\downarrow\uparrow\uparrow\uparrow\uparrow\downarrow\downarrow\downarrow\downarrow}$, respectively. For all results, we set $J=1$, $\gamma=0.5$, and $\delta t=0.1$, and observe very good agreement between exact numerics and hardware quantum simulation that all showcase qualitatively similar non-Hermitian skin pumping towards site $i=0$.}}
	\label{fig:dynamics}
\end{figure}
%\subsection{Impact of gate noise}
%\begin{figure}
%	\centering
%	\includegraphics[width=1\linewidth]{dynamics2}
%	\caption{}
%	\label{fig:dynamics2}
%\end{figure}

\rz{The comparison between the details of processors shown in FIG.~\ref{fig:ibmq_device_supp} and FIG.~\ref{fig:ibm} indicates a significant hardware improvement where the gate errors have been significantly reduced by IBM Q. Given this enhanced performance, we conducted additional simulations on the improved hardware ``ibm\_brisbane". Specifically, we simulate the dynamics of the non-interacting Hatano-Nelson model (Eq.~\ref{HNsupp}) for different physical system sizes. These hardware simulations are shown in FIG.~\ref{fig:dynamics}. In FIGs.~\ref{fig:dynamics}a,b,c, where the system sizes are 7, 9, and 11 qubits respectively, our hardware simulations on IBM Q show excellent agreement with the theoretical results. Although the dynamical density profiles for 13 qubits in FIGs.~\ref{fig:dynamics}d are more affected by the hardware noise, the skin accumulation can still be clearly observed. This is because while the improved hardware performs well for smaller systems, simulations of larger systems are still impacted by the noise due to a larger number of gates present in the quantum circuit. Neverthelss, the error itself is still not enough to affect the conclusive observation of non-Hermitian skin pumping in our systems considered.}
	
%\newpage
\flushbottom
%%%%%%%%%%%%%%%%%%%%%%  References %%%%%%%%%%%%%%%%%%%%%%
%\bibliographystyle{apsrev4-1}
%\bibliographystyle{plain}
%\bibliography{references}
%\subsection*{\normalsize Supplementary references}
%\bibliography{ref}

\end{document}